\documentclass[fleqn,usenatbib]{mnras}

\usepackage{newtxtext,newtxmath}

\usepackage[T1]{fontenc}
\usepackage{ae,aecompl}

\usepackage{graphicx}	
\usepackage{amsmath}	
\usepackage{subfigure}
\usepackage{multirow}

\usepackage{tikz}
\usetikzlibrary{arrows.meta}
\tikzset{>={Latex[width = 1mm,length = 2mm]},
    base/.style = {rectangle, draw = blue, thick, minimum width = 8cm, minimum height = 1cm, text centered, font = \sffamily}, 
    repeat1/.style = {base, rounded corners, draw = red, minimum width = 7cm}, 
    repeat2/.style = {repeat1, draw = orange, minimum width = 6cm}
}

\def\code#1{{\tt \textsc{#1}}}

\usepackage{stfloats}

\title[Hierarchical Clusters in Haloes]{The Hierarchical Structure of Galactic Haloes: 
\\
Classification and characterisation with \textsc{Halo-OPTICS}}

\author[Oliver et al.]{
William H. Oliver$^{1}$\thanks{E-mail: william.oliver@sydney.edu.au},
Pascal J. Elahi$^{2,3}$,
Geraint F. Lewis$^{1}$, and
Chris Power$^{2,3}$
\\
$^{1}$Sydney Institute for Astronomy, School of Physics A28, The University of Sydney, NSW, 2006, Australia\\
$^{2}$International Centre for Radio Astronomy Research, University of Western Australia, 35 Stirling Highway, Crawley, WA, 6009, Australia\\
$^{3}$ARC Centre of Excellence for All Sky Astrophysics in 3 Dimensions (ASTRO 3D), University of Western Australia, 35 Stirling Highway, Crawley, WA 6009,\\Australia 
}

\date{Accepted XXX. Received YYY; in original form ZZZ}

\pubyear{2020}

\begin{document}
\label{firstpage}
\pagerange{\pageref{firstpage}--\pageref{lastpage}}
\maketitle

\begin{abstract}
We build upon \textbf{O}rdering \textbf{P}oints \textbf{T}o \textbf{I}dentify \textbf{C}lustering \textbf{S}tructure (\code{OPTICS}), a hierarchical clustering algorithm well-known to be a robust data-miner, in order to produce \code{Halo-OPTICS}, an algorithm designed for the automatic detection and extraction of all meaningful clusters between any two arbitrary sizes. We then apply \code{Halo-OPTICS} to the 3D spatial positions of halo particles within four separate synthetic Milky Way type galaxies, classifying the stellar and dark matter structural hierarchies.
Through visualisation of the \code{Halo-OPTICS} output, we compare its structure identification to the state-of-the-art galaxy/(sub)halo finder \code{VELOCIraptor}, finding excellent agreement even though \code{Halo-OPTICS} does not consider kinematic information in this current implementation.
%
%
We conclude that \code{Halo-OPTICS} is a robust hierarchical halo finder, although its determination of lower spatial-density features such as the tails of streams could be improved with the inclusion of extra localised information such as particle kinematics and stellar metallicity into its distance metric.
\end{abstract}




\section{Introduction} \label{sec:introduction}
A primary prediction of hierarchical galaxy formation in the Cold Dark Matter (CDM) cosmological model is that galaxies should be surrounded by numerous low-mass satellites as a result of historical and  ongoing accretion \citep{White1978, Kauffmann1993, Ghigna1998}. Simulations of galaxy formation under this regime have predicted that galaxies of a size similar to the Milky Way (MW) could harbour $300$--$500$ satellites at least as massive as $\sim10^8\ \rm{M_\odot}$ at $z=0$ \citep{Klypin1999, Moore1999, Reed2005, Tollerud2008, Springel2008, Ishiyama2013}. Observations within and around the MW have identified $\sim60$ satellites of the same size; refer to Tables A1 and A2 from \citet{Newton2018} and more recently \citet{Koposov2018, Mau2019, Torrealba2019, Homma2019} for catalogues of these. This has become known as the missing satellite problem.

Studies have shown that by suppressing the small-scale power spectrum \citep{Kamionkowski2000, Zentner2003}, by considering warm dark matter as an alternative model \citep{Colin2000, Bode2001}, or by enforcing that dark matter emerged from the late decay of a non-relativistic particle \citep{Strigari2007}, the inconsistency is brought to within a reasonable margin of error as a result of a reduced theoretical number of satellites (as well as the slowly increasing number of the observed). In conjunction with these cosmological solutions, the observations of the faintest MW satellites have also suggested that the discrepancy is smaller than previously thought and that smaller dark matter haloes are far less efficient at forming stellar populations than their more massive counterparts \citep{Bullock2000, Somerville2002, Benson2002, Ricotti2005, Moore2006}. The implication is that the observed mass-scale of MW dwarfs may be an artefact of detection bias. By considering more complex baryonic physics such as re-ionisation and supernovae feedback, the most recent cosmological N-body simulations suggest that the missing satellite problem may no longer be an obstacle for the CDM model \citep{Brooks2014, Sawala2015, Dutton2016, Sawala2016, Wetzel2016, Zhu2016, Kim2018, Fielder2019}.

Even though the tension of CDM with the number of satellites has relaxed, the search for satellites in the Local Group is ongoing. In simulations, the abundance of subhaloes has been shown to be dependent on the mass of the subhalo such that ${\rm d}n_{\rm sub}/{\rm d}M_{\rm sub} \propto M_{\rm sub}^{-\alpha}$ with $\alpha \approx 1.9$ \citep{Gao2004, Reed2005, Diemand2007, Springel2008, Angulo2009, GarrisonKimmel2014, Xie2015, RodriguezPuebla2016, Elahi2018}, where $n_{\rm sub}$ is the number of subhaloes with mass greater than $M_{\rm sub}$ -- the individual subhalo masses. This relation is appropriately descriptive of subhaloes whose dark matter hosts have masses within the range of $10^{12}$--$10^{15} \rm{M_\odot}$, although this range has a lower bound set by the each simulation's particle mass resolution; however, it is expected to hold true for host halo masses much smaller than this. Observationally, this power law appears to hold for subhaloes with luminosity down to $10^{8}\ \rm{L_\odot}$ -- approximately the luminosity of the brightest satellites of the MW and M31 -- although it is not consistent for those satellites whose luminosity is below that limit \citep{Tollerud2014}. This could again support the notion that currently our satellite detection capabilities fail to easily detect those ultra-faint satellites \citep{Koposov2008, Tollerud2008, Walsh2008, Bullock2010, Sesar2014} that are hosted by the aforementioned dark matter subhaloes responsible for suppressing the star formation within them \citep{Wolf2010, Martinez2011}.

Surveys have exposed a considerable amount of substructure surrounding the MW \citep[e.g.][]{LAMOST, Pristine, SDSSIV, GaiaDR2} and other nearby galaxies such as M31 \citep{McConnachie2009, McConnachie2018}. We see that while this structure may not be directly associated with specific satellites of these galaxies, it is still intimately linked to them having arisen from the satellite-host tidal interactions. Characterising the observed structure quantitatively will complement studies of satellite galaxies. As such an important component of subhalo analysis is the initial determination of such objects. There are many algorithms used to ascertain clustered data from data sets. Data miners such as \code{K-Means} \citep{Lloyd1982}, \code{Mean-Shift} \citep{Fukunaga1975}, and \code{DBSCAN} \citep{Ester1996} work well for certain cluster shapes although they will not indicate the hierarchy of clusters present in a data set. Astrophysical clustering algorithms like \code{SUBFIND} \citep{Springel2001}, Robust Overdensity Calculation using K-Space Topologically Adaptive Refinement \citep[{\code{ROCKSTAR}};][]{Behroozi2012}, Amiga Halo Finder \citep[{\code{AHF}};][]{Knollmann2009}, \code{VELOCIraptor} \citep{Elahi2019}, and others (refer to \citet{Knebe2011} for a comparison of these and 14 others) mostly rely on the Spherical-Overdensity method \citep[{\code{SO}};][]{Press1974}, Friends-Of-Friends \citep[{\code{FOF}};][]{Davis1985}, or some iterative combination of the two.

The \code{SO} method aims at identifying the density peaks enclosed within some dense region of N nearest-neighbours. Following this, spherical surfaces expand about each peak until a specified overdensity is achieved within it whilst iteratively adapting the centre of the sphere to the new centroid of the enclosed particles. The biggest downfall of the \code{SO} method is that it fails to detect clusters below the specified overdensity threshold. Contrarily, the \code{FOF} algorithm endeavours to link together those particles that are physically close to each other and then subsequently computes the centroid of this particle composition. Neither the \code{SO} method nor the \code{FOF} algorithm are inherently hierarchical unless used iteratively on the findings of their previous applications with a larger overdensity or smaller linking length, respectively. Only then can these algorithms differentiate between clusters of two different densities within the same hierarchy. A novel and complementary algorithm to the traditional structure finders which is hierarchical in this sense is the \textbf{O}rdering \textbf{P}oints \textbf{T}o \textbf{I}dentify \textbf{C}lustering \textbf{S}tructure (\code{OPTICS}) algorithm \citep{Ankerst1999}. \code{OPTICS} can be readily applied to observational data sets in ways that some traditional structure finders cannot since it does not require estimates of the gravitational potential.

We build upon \code{OPTICS} to develop \code{Halo-OPTICS} and apply it to the synthetic galactic haloes generated by \citet{Power2016} at redshift zero. We first summarise the details of this data set and outline our choice of physical quantities from these synthetic haloes in Sec. \ref{subsec:synthetichaloes}. We then summarise the \code{OPTICS} algorithm in detail in Sec. \ref{subsec:optics}. Next, we present \code{Halo-OPTICS}; by first justifying our choice of the \code{OPTICS} hyperparameters $\epsilon$ and $N_{\rm{min}}$ in Sec. \ref{subsec:opticsparams}, then defining our automatic cluster extraction technique in Sec. \ref{subsec:structuredetection}. We conduct parameter optimisation tests in Sec. \ref{subsec:purityandrecovery} and inform the reader about the nature of the \code{Halo-OPTICS} hierarchy in Sec. \ref{subsec:hierarchyexplanation}. We then present our findings; by visualising the \code{Halo-OPTICS} output in Sec. \ref{subsec:opticsoutput}, comparing \code{Halo-OPTICS} with \code{VELOCIraptor} in Sec. \ref{subsec:comparison}, and by analysing the galactic hierarchy returned by \code{Halo-OPTICS} in Sec. \ref{subsec:highresolutionzone}. Later in Sec. \ref{sec:discussion}, we discuss these results and the implications of providing \code{Halo-OPTICS} with extra localised information. Finally, we make our conclusions and express our intent for future works in Sec. \ref{sec:conclusions}.

\section{Background} \label{sec:background}
\subsection{Synthetic Haloes} \label{subsec:synthetichaloes}
For our synthetic halo data we use those produced by \citet{Power2016} at redshift zero. These haloes are drawn from a set of cosmological zoom simulations. The parent simulation \citep[run with {\tt GADGET-2};][]{Springel2005} is a $\Lambda$CDM N-body simulation conducted in a $50\ \rm{Mpc/h}$ cube with $256^3$ particles. The total matter, baryon, and dark energy density parameters are $\Omega_m = 0.275$, $\Omega_b = 0.0458$, and $\Omega_\Lambda = 0.725$ respectively, and the dimensionless Hubble constant is ${\rm h} = 0.702$. The power spectrum normalisation is $\sigma_8 = 0.816$, and the primordial spectral index is $n_s = 0.968$. At $z = 0$, the \code{FOF} algorithm was used to select MW type haloes with $M_{200} \approx 2\times 10^{12}\ \rm{M_\odot/h}$ that reside in low-density (void) regions, which were identified with the V-web algorithm of \citet{Hoffman2012}. These galaxies were then re-simulated with a version of {\tt GADGET-3}, as discussed in \citet{Power2016}, from $z = 99$ to $z = 0$ using all particles contained within a radius of $5R_{200}$. The re-simulations include the baryonic physics of cooling, star formation, supernovae feedback, but do not include any chemical evolution. More details on these simulations are found in \citet{Power2016}. The stellar particle mass and dark matter particle mass in these re-simulated galaxies are $M_s \approx 10^6\ \rm{M_\odot/h}$ and $M_d \approx 5\times10^6\ \rm{M_\odot/h}$ respectively.

\begin{figure*}
    \includegraphics[trim={9.5mm 10mm 17.5mm 23mm}, clip, width=0.7\textwidth]{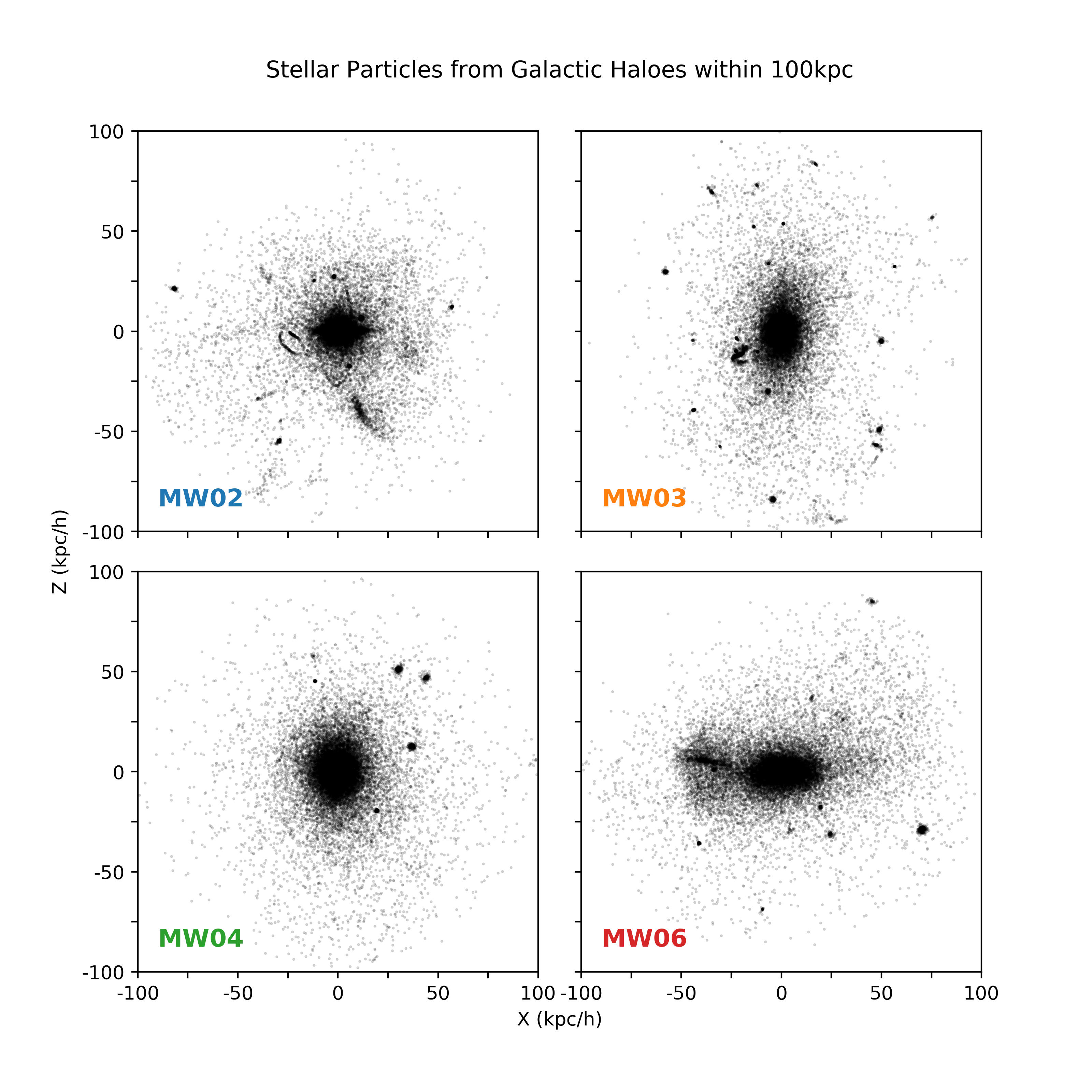}
    \vspace{-0.15cm}
    \caption{A 2D projection of each of the synthetic haloes within $100\ {\rm kpc/h}$ of their barycentre. Each of the panels are marked in their lower left according to the galaxy they contain. The colour scheme of these markings correspond to those used to distinguish the galaxies in Fig. \ref{fig:hierarchy}. It can be seen that each MW type galaxy contains a dense central region with an abundance of substructure surrounding it.}
    \label{fig:MultiGalaxyView}
\end{figure*}

To appropriately consider the structures within stellar haloes we use an open-source python package for data analysis, namely yt \citep{yt}, to separately read the 3D positions as well as the masses for all stellar and dark matter particles present in each of the $\Lambda$CDM synthetic haloes at $z = 0$. To be unambiguous, we only use the 3D positions of these particles to identify the presence of clustering in the data sets.  Fig. \ref{fig:MultiGalaxyView} provides a visualisation of each of the synthetic haloes within $100\ {\rm kpc/h}$ of their barycentre. The barycentres are determined using the {\it shrinking spheres} method outlined in \citet{Power2003}. Fig. \ref{fig:MultiGalaxyView} indicates that there is an abundance of hierarchical substructure present within each galactic halo.

\subsection{Ordering Points To Identify Clustering Structure} \label{subsec:optics}
The \code{OPTICS} algorithm is a robust tool for hierarchically identifying density-based structure in any n-dimensional data set for which a distance metric can be defined. It has been used across various fields to; quantify human behaviour and mobility patterns \citep{Zheng2008}, characterise the genomic diversity in wheat \citep{Wang2014}, optimise the distribution of urban energy supply systems \citep{Marquant2017} and more. For data sets containing variables with incompatible units, the distance metric can be difficult to construct. However, given a data set of spatial coordinates, the choice of a distance metric is obvious, namely the Euclidean distance. This certainty makes the application of \code{OPTICS} to the physical clustering of particles in 3D space very powerful and robust. Despite this, \code{OPTICS} has only been applied in an astrophysical context a handful of times \citep[e.g.][]{Fuentes2017, McConnachie2018, Canovas2019, Massaro2019}\footnote{\citet{Fuentes2017} examine the suitability of \code{OPTICS} to be applied to large scale data sets by testing its performance when applied to a simulated astrophysical data set. \citet{McConnachie2018, Canovas2019, Massaro2019} apply \code{OPTICS} to observational data sets to identify both new members of existing clusters and new clusters entirely. We build upon these works in order to produce \code{Halo-OPTICS} by; standardising the approach under which \code{OPTICS} should be applied to astrophysical data sets, establishing a cluster extraction method that appropriately identifies a full hierarchy of astrophysical clusters from the \code{OPTICS} output, and verifying its performance.}.

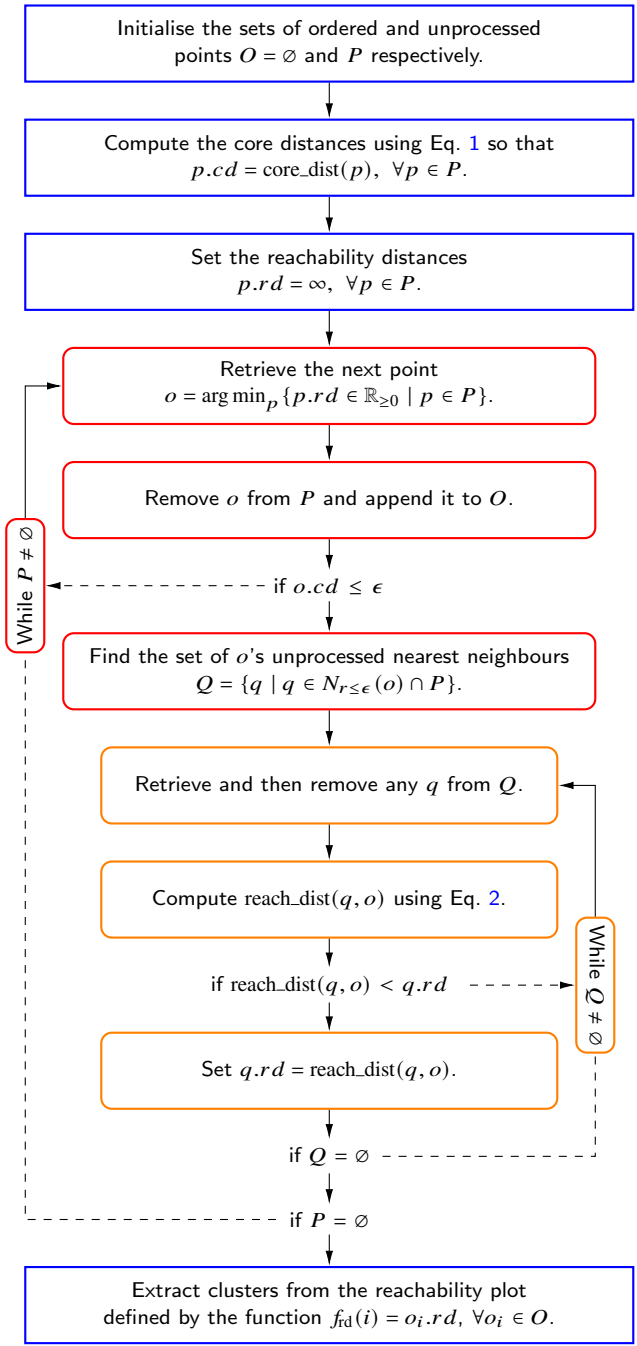
\begin{figure}
\begin{center}
\begin{tikzpicture}[node distance=1.5cm, every node/.style={fill=white, text centered, font=\sffamily}], align=center]
    \node[align = center] (setup)[base]{Initialise the sets of ordered and unprocessed\\points $O = \varnothing$ and $P$ respectively.};
    
    \node[align = center] (coredist)[base, below of=setup]{Compute the core distances using Eq. \ref{eq:coredistance} so that\\$p.cd = {\rm core\_dist}(p),\ \forall p \in P$.};
    
    \node[align = center] (reachdist)[base, below of=coredist]{Set the reachability distances\\$p.rd = \infty,\ \forall p \in P$.};
    
    \node[align = center] (nextpoint)[repeat1, below of=reachdist]{Retrieve the next point\\$o = \operatorname*{arg\,min}_p \{p.rd \in \mathbb{R}_{\geq 0}\mid p \in P\}$.};
    
    \node[align = center] (processpoint)[repeat1, below of=nextpoint]{Remove $o$ from $P$ and append it to $O$.};
    
    \node(ifcd)[text width = 1.5cm, below of=processpoint, yshift=0.375cm]{if $o.cd$ $\leq$ $\epsilon$};
    
    \node[align = center] (nearestneighbours)[repeat1, below of=ifcd, yshift=0.375cm]{Find the set of $o$'s unprocessed nearest neighbours\\$Q = \{q \mid q \in N_{r \leq \epsilon}(o) \cap P\}$.};
    
    \node (chooseq)[repeat2, below of=nearestneighbours]{Retrieve and then remove any $q$ from $Q$.};
    
    \node (computerd)[repeat2, below of=chooseq]{Compute ${\rm reach\_dist}(q, o)$ using Eq. \ref{eq:reachabilitydistance}.};
    
    \node (ifrd)[text width = 3.5cm, below of=computerd, yshift=0.375cm]{if ${\rm reach\_dist}(q, o)$ $<$ $q.rd$};
    
    \node (setrd)[repeat2, below of=ifrd, yshift=0.375cm]{Set $q.rd = {\rm reach\_dist}(q, o)$.};
    
    \node (unprocessed)[repeat1, left of=nearestneighbours, xshift=-2.5cm, yshift=1.125cm, rotate=90, minimum width = 1cm, minimum height = 0.5cm]{While $P \neq \varnothing$};   
    
    \node (rdforNN)[repeat2, right of=setrd, xshift=2cm, yshift=1.125cm, rotate=270, minimum width = 1cm, minimum height = 0.5cm]{While $Q \neq \varnothing$};  
    
    \node (ifqempty)[below of=setrd, text width = 1.2cm, yshift=0.375cm]{if $Q$ $=$ $\varnothing$};
    
    \node (ifpempty)[below of=ifqempty, text width = 1.2cm, yshift=0.67cm]{if $P$ $=$ $\varnothing$};
    
    \node[align = center] (clusterextraction)[base, below of=ifpempty, yshift=0.375cm]{Extract clusters from the reachability plot\\defined by the function $f_{\rm rd}(i) = o_i.rd$, $\forall o_i \in O$.};
    
    \draw[->](setup) -- (coredist);
    \draw[->](coredist) -- (reachdist);
    \draw[->](reachdist) -- (nextpoint);
    \draw[->](nextpoint) -- (processpoint);
    \draw[->](processpoint) -- (ifcd);
    \draw[->](ifcd) -- (nearestneighbours);
    \draw[dashed, ->](ifcd) -- (unprocessed);
    \draw[->](nearestneighbours) -- (chooseq);
    \draw[->](chooseq) -- (computerd);
    \draw[->](computerd) -- (ifrd);
    \draw[->](ifrd) -- (setrd);
    \draw[dashed, ->](ifrd) -- (rdforNN);
    \draw[->](setrd) -- (ifqempty);
    \draw[->](ifqempty) -- (ifpempty);
    \draw[->](ifpempty) -- (clusterextraction);
    \draw[dashed](ifpempty) -| (unprocessed);
    \draw[<-](nextpoint) -| (unprocessed);
    \draw[dashed](ifqempty) -| (rdforNN);
    \draw[<-](chooseq) -| (rdforNN);
\end{tikzpicture}
\end{center}
\vspace{-8.7314pt}
\caption{The \code{OPTICS} activity chart with nodes outlined in blue, red, and orange to indicate that they are one-time operations, part of the outer while loop, or part of the inner while loop respectively. The solid and dashed lines indicate the paths to be taken if a condition is or is not satisfied respectively. Paths that are not conditional are also shown as solid lines. Paths with arrow heads are uni-directional whereas those without arrow heads may be traversed in either direction depending on the current context -- i.e. the process moves towards satisfied conditions. Traditionally the core distances are computed for each point individually after they are appended to $O$, however computing them collectively speeds up the process and also allows this step to be run in parallel without changing the reachability plot. The inner while loop may also be run in parallel for any given $o$, though the outer while loop cannot be parallelised due to the sequential data access order. It should be noted that there are no additional constraints when retrieving the next point to be ordered into $O$, if multiple points have equal reachability distances then the next point is chosen randomly from them. It is also due to this that the first ordered point is simply a random element of $P$.}
\label{fig:opticsprocess}
\end{figure}

For each point in a data set, \code{OPTICS} calculates a measure of the local density surrounding that point called a reachability distance -- see Eq. \ref{eq:reachabilitydistance} below. \code{OPTICS} also concurrently creates an ordered list of all points in the data set, such that any point with an ordered index of $i$ is the most {\it reachable} previously unordered point to all points with an ordered index less than $i$. The visualisation of the output of the \code{OPTICS} algorithm is the reachability plot -- the reachability distances as a function of the ordered index for all points in the data set. Fig. \ref{fig:opticsprocess} depicts the way in which \code{OPTICS} achieves this. The reachability distance is an inverse measure of the local density surrounding each point and as such, clusters in the data set present themselves as valleys in the reachability plot. Refer to Sec. \ref{subsec:structuredetection} and Fig. \ref{fig:ExtractionProcess} for details on the algorithm we use to extract meaningful clusters from the reachability plot.

\begin{figure}
    \includegraphics[trim={24.5mm 8mm 14.5mm 34.5mm}, clip, width=\columnwidth]{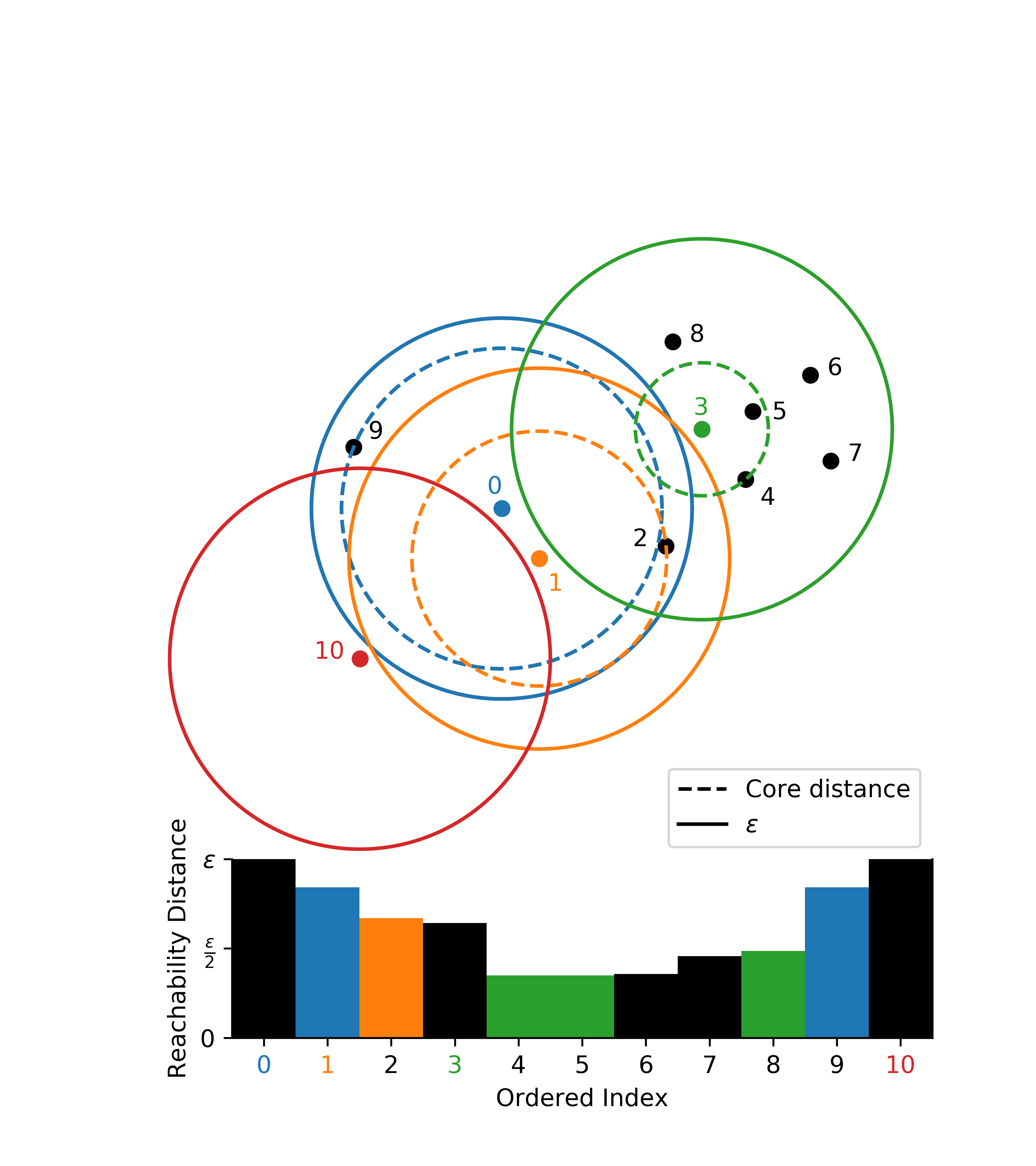}
    \vspace{-0.5cm}
    \caption{A 2D toy example of the \code{OPTICS} algorithmic process as well as the corresponding reachability plot output. This example is conducted using $N_{\rm{min}}=3$ with an arbitrarily scaled $\epsilon$ parameter, the corresponding core distances and nearest-neighbour search radius are also shown for the ordered indices $0$, $1$, $3$, and $10$. The colour of each coloured bar in the reachability plot corresponds to the colour of the point that produced that bar's reachability distance value. Notice that the points with ordered indices $0$ and $10$ actually have reachability distances of infinity (although it is not shown here) as a result of never having previously (at the time they are appended to the ordered list) been a part of any other core-point's unprocessed nearest neighbourhood -- refer to Fig. \ref{fig:opticsprocess} and the main text of Sec. \ref{subsec:optics} for more details on the process.}
    \label{fig:OpticsExplanation}
\end{figure}

The power of \code{OPTICS} is in part owed to the minimal input required on the user's behalf. \code{OPTICS} only requires the user to provide two parameters, $\epsilon$ and $N_{\rm{min}}$. These parameters are chosen according to the data set and are robust enough that small changes in their choice do not strongly affect the reachability plot nor the determination of any clusters present in the data\footnote{The \code{OPTICS} output can be greatly affected if the fractional change in $N_{\rm min}$ is large even if the change in $N_{\rm min}$ itself is small.}. Refer to Sec. \ref{subsec:opticsparams} for details regarding our choice of these parameters.

\begin{enumerate}
  \item The parameter $\epsilon$ is the radius for which a nearest-neighbour radius query is performed for each point in the data set, and consequently is also the largest possible reachability distance for any point. An appropriate choice of $\epsilon$ is made through a consideration of the trade-off between the least dense structures that the user wishes to detect, as well as the run-time of the algorithm.
  \item The parameter $N_{\rm{min}}$ is the minimum number of points that a structure must contain such that it can be detected as a cluster. This parameter is also fundamental in the calculation of the reachability distance. Increasing $N_{\rm{min}}$ reduces noise in the reachability plot, but limits the smallest possible structures determinable to clusters containing at least $N_{\rm{min}}$ points.
\end{enumerate}

For a given point $o$ in the data set, if at least $N_{\rm{min}}$ points are returned from its nearest-neighbour radius query -- i.e. $|N_{r \leq \epsilon}(o)| \geq N_{\rm{min}}$ \footnote{Note that by convention a point $o$ is included amongst its own nearest-neighbour search and hence the $N_{\rm{min}}$ nearest-neighbours include $o$ as well.}, then that point is labelled as a core-point. Every point is assigned a core distance such that
\begin{equation}\label{eq:coredistance}
    {\rm core\_dist}(o) = ||o - p||_2,
\end{equation}
where $p$ is $o$'s ${N_{\rm min}}^{\rm th}$-most nearest-neighbour. It then follows that every core-point will have a core distance less than or equal to $\epsilon$. Given that $o$ is a core-point, it will at some stage of the algorithm be used to ascertain the reachability distance for each of its currently (at that stage) unprocessed nearest-neighbours, $q$, within a radius of $\epsilon$ from $o$ -- i.e. $q \in N_{r \leq \epsilon}(o) \cap P$ where $P$ are those {\it currently} unprocessed points. The reachability distance for each of these nearest-neighbours with respect to $o$ is
\begin{equation}\label{eq:reachabilitydistance}
    {\rm reach\_dist}(q, o) = {\rm max}({\rm core\_dist}(o), ||q - o||_2).
\end{equation}
This ensures that the closest $N_{\rm{min}}$ points to $o$ have a reachability distance with respect to $o$ equal to the core distance of $o$, while all other nearest-neighbours of $o$ have a reachability distance with respect to $o$ equal to their Euclidean distance from $o$. The reachability distance with respect to $o$ of any given unprocessed nearest-neighbour $q$ is assigned to $q$ if it is smaller than $q$'s {\it currently} assigned reachability distance.

For clarity, each point in the data set is initialised with a reachability distance of infinity and the ordered list is determined by iteratively appending to it; the point with the smallest reachability distance. For each iteration, the above set of reachability distances with respect to the current point $o$ are calculated and used to adjust the reachability distances in the data set as is described above and as in Fig. \ref{fig:opticsprocess}.

This process of adjusting the reachability distance ensures that the reachability plot remains smooth, contains less noise than it otherwise would, and ultimately gives a reliable representation of the density of any structures present within the data set. Furthermore, it is an effective process for limiting the algorithm's {\it knowledge} of local densities at any particular iteration to those points that have been ordered up until that iteration, whilst concurrently seeking out the regions of highest density from them. This process makes the reachability distance non-deterministic, as it depends upon the ordered list. As such, the final reachability distance of a point, $q$, is always $\geq$ the smallest core distance of the points in the set of $q$'s $N_{\rm min}$ reverse nearest-neighbours -- which is defined to be the set of all points in the data whose $N_{\rm min}$ nearest-neighbours contain $q$. Consequentially, and although not intended for this purpose, the reachability distance of a point (or rather the density within the volume of the n-sphere it encompasses) can only be interpreted as an approximate density estimator that has been found at the resolution of $N_{\rm min}$ nearest neighbours. Density estimators commonly used in numerical cosmology such as the SPH \citep{Monaghan1992} and Voronoi \citep{Voronoi1908, vanDeWaygaert1994, Okabe2016} estimators are deterministic and do provide a unique measure of density for each point - qualities that the reachability distance does not possess. Ultimately, the reachability distance and the process under which it is created provides not only an approximate measure of local density but more importantly a means for ordering the points of a data set and thereby reducing the n-dimensionality of the clustering structure to a 2D representation of it. Fig. \ref{fig:OpticsExplanation} shows a demonstration of the \code{OPTICS} process for a 2D toy data set. Here each point has been marked corresponding to its ordered index. Furthermore, the core distances and commonly shared $\epsilon$ parameter have been marked and uniquely coloured for the ordered indices $0$, $1$, $3$, and $10$. The figure illustrates that the more spatially clustered points have been consecutively ordered in the reachability plot and have been assigned smaller reachability distances than the other points.

One of the major drawbacks to \code{OPTICS} is that it is computationally demanding, particularly for large data sets. \citet{Ankerst1999} report a constant factor increase in run-time of $1.6$ when compared to its predecessor \code{DBSCAN} \citep{Ester1996}. The main difference between the two being that \code{DBSCAN} returns one level of clustering, that is, all lists of points that are densely connected through core-point neighbourhoods. Whereas \code{OPTICS} extends the rigidity of a point either being part of a cluster or not, to a measure of {\it how much is a point a part of a cluster} through the means of the reachability distance. The worst-case time-complexity for \code{OPTICS} is $\mathcal{O}(n^2)$, although in general the time-complexity is $\mathcal{O}(n\times r_\epsilon)$ -- where $r_\epsilon$ is the average run-time of the nearest-neighbour radius queries. Choosing fast nearest-neighbour search algorithms, such as scikit-learn's \code{KDTree} algorithm \citep{Bentley1975, scikit-learn} ($\mathcal{O}(r_\epsilon) \to \mathcal{O}(log(n))$), as well as taking small values for $\epsilon$, help to reduce the run-time of a nearest-neighbour radius query. Other improvements of the total run-time can be made through seed list optimisation, partial sorting, and parallelisation \citep{Fuentes2017}, although the parallelisation of \code{OPTICS} is notoriously difficult due to its strongly sequential data access order, and typically the algorithm must be altered \citep{Patwary2013}. Another property of \code{OPTICS} is that it does not naturally identify the clusters present within the data since it only returns a measure of local density about each point. However as is explained below in Sec. \ref{subsec:structuredetection}, this is also its most advantageous quality as it allows the user to be specific about the density, hierarchy level distinction, and point inclusion criteria that they wish to use to define a cluster.

\section{\textsc{Halo-OPTICS}: A Hierarchical Galaxy/(Sub)Halo Finder} \label{sec:halooptics}
\subsection{Choosing Appropriate \textsc{OPTICS} Hyperparameters} \label{subsec:opticsparams}
As is mentioned in Sec. \ref{subsec:optics}, the choices of $\epsilon$ and $N_{\rm{min}}$ are determined through a consideration of the minimum size and density of structures the user wishes to detect, as well as the run-time constraints the user wishes to adhere to. In order to provide a substantially high degree of resolution for the identifiable clusters, we choose to set $N_{\rm{min}} = 20$ -- a common choice for the minimum size of meaningful clusters in substructure finders. The choice to set $N_{\rm{min}}$ as a constant between each application to the galactic haloes is implemented so that the minimum possible mass of the clusters remains roughly equal between them (and only differs due to the small differences between particle masses in these simulations). The lower mass limit of clusters is $\approx 2\times10^7\ \rm{M_\odot/h}$ for stellar clusters and $\approx 1\times10^8\ \rm{M_\odot/h}$ for dark matter clusters. This ensures that we can meaningfully compare the clustering between different particle types and different haloes. Having such a small value of $N_{\rm{min}}$ does however introduce more noise to the reachability plot than when compared to that of larger $N_{\rm{min}}$ values. This extra noise makes meaningful cluster extraction more involved and less obvious, and so we have constructed our own algorithm for automatic cluster detection which we present in Sec. \ref{subsec:structuredetection}.

Making the choice for $\epsilon$ is a little more difficult as this essentially specifies the size and extent of the least dense structures. The \code{FOF} analogue for this parameter is the linking length which, in cosmological simulations and given a halo with virial overdensity $\Delta$ that contains $N_\Delta$ particles within a radius of $R_\Delta$, may be chosen using $l_x = (2\pi/N_\Delta)^{1/3}R_\Delta$ \citep{Elahi2011}. To extrapolate this to the $\epsilon$ parameter we need to account for the fundamental difference between the algorithms. A point will only be assigned a reachability distance if it has previously been included in a now-ordered core-point's set of unprocessed nearest neighbours within a radius of $\epsilon$. Therefore, the least dense core such a point can be a part of is the core that surrounds the core-point that has exactly $N_{\rm{min}}$ nearest neighbours that extend out to exactly a radius of $\epsilon$. This can be leveraged to find the factor by which $\epsilon$ must be larger than the \code{FOF} linking length to encompass the same overdensity\footnote{Due to the fundamental differences between \code{OPTICS} and \code{FOF} there will not, in general, be a value for $\epsilon$ that produces precisely the same grouping of points as \code{FOF} would by using a particular linking length $l_x$. This is because \code{OPTICS} does not care about intra-core structure while \code{FOF} does, i.e. \code{OPTICS} is less susceptible to point-point noise than \code{FOF} is. It should also be noted that constructing $\epsilon$ from $l_x$ in this way is an approximation for finding haloes that enclose a specified overdensity. As detailed in the study by \citet{More2011}, there does not exist a unique linking length that encloses a specified overdensity for all \code{FOF} haloes. It is shown therein that the resultant enclosed overdensity of \code{FOF} haloes is not only dependent on the number of particles in the halo but also the concentration of the density peak. As such, the enclosed overdensity of \code{Halo-OPTICS} haloes is also subject to this ambiguity. However, since the mapping between a \code{FOF} halo and a \code{Halo-OPTICS} halo is not exact either, we do not find it necessary to conduct a more thorough determination of $l_x$ for the purpose of computing $\epsilon$.}.

For each application of \code{Halo-OPTICS}, we first compute the corresponding \code{FOF} linking length (as detailed above) and then seek to multiply this by the mean of the end-to-end distance of a chain of $N_{\rm{min}}$ points, each separated by unit length and each sampled from a directionally uniform probability distribution. Even though we only use $N_{\rm{min}} = 20$ in this work, the following describes how we prepare the code for use with all possible values of $N_{\rm{min}}$. Since there is no analytical solution to the mean of this distance distribution in terms of elementary functions, we run a series of Monte Carlo simulations for $N_{\rm{min}} = 3$ to $20$ that compute it and store them as hard coded variables within the code. The case when $N_{\rm{min}} = 2$ is trivial since this is identical to the \code{FOF} case and $\epsilon = l_x$. For larger values of $N_{\rm{min}}$ we can base our approximation off the root-mean-square distance since it does have an analytical formula and is given by $d_{\rm RMS}(N_{\rm{min}}) = \sqrt{N_{\rm{min}} - 1}$. The root-mean-square of the end-to-end distance of the chain approaches a constant value of $\sim1.084$ times the Monte Carlo simulated mean of the end-to-end distance of the chain. Therefore, a good approximation of the mean of the end-to-end distance of the chain for $N_{\rm{min}} > 20$ is given by $d_{\rm RMS}(N_{\rm{min}})/1.084$. For a set of $10$ roughly log-spaced integers between $N_{\rm{min}} = 21$ and $1000$, this approximation is always within $0.2\%$ of the mean found from Monte Carlo simulations. All Monte Carlo simulations used here calculate $10^7$ separate chains from which the end-to-end distances and resultant means are found. This calculation of $\epsilon$ from $N_{\rm{min}}$ and the \code{FOF} linking length, $l_x$, effectively switches the \code{OPTICS} input parameter $\epsilon$, to the physically motivated \code{Halo-OPTICS} input parameter $\Delta$, which we choose as $\Delta = 200$ (times the critical density of the universe). This also means that the root haloes (refer to \ref{subsec:structuredetection} for a breakdown of this terminology) found by \code{Halo-OPTICS} encompass a similar overdensity to those that would be found by an \code{FOF}-based code.
 
\subsection{Extracting Clusters from \textsc{OPTICS}} \label{subsec:structuredetection}
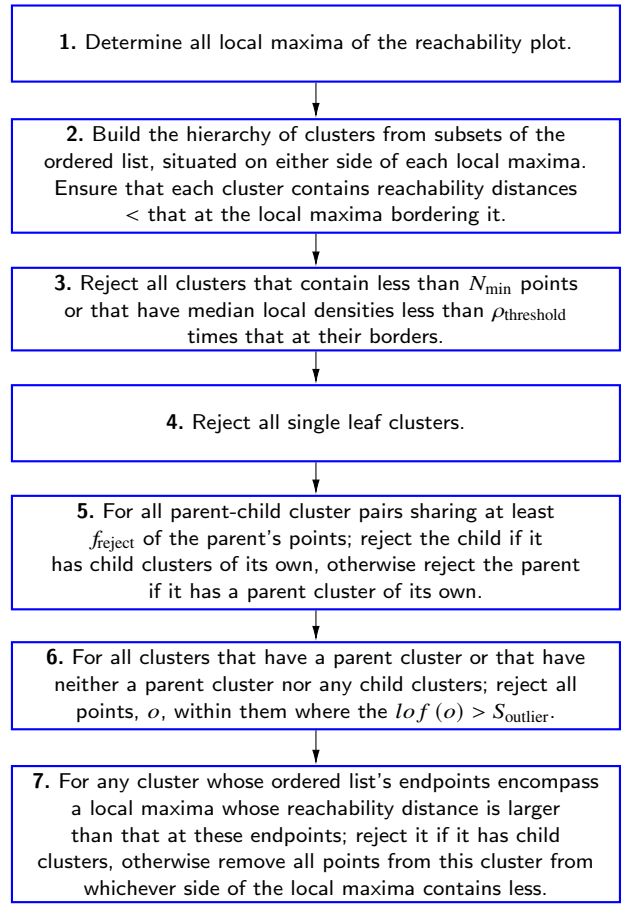
\begin{figure}
\begin{center}
\begin{tikzpicture}[node distance=1.5cm, every node/.style={fill=white, text centered, font=\sffamily}], align=center]
    \node[align = center] (step1)[base]{\textbf{1.} Determine all local maxima of the reachability plot.};
    
    \node[align = center] (step2)[base, below of=step1, yshift = -2.4mm]{\textbf{2.} Build the hierarchy of clusters from subsets of the\\ordered list, situated on either side of each local maxima.\\Ensure that each cluster contains reachability distances\\$<$ that at the local maxima bordering it.};
    
    \node[align = center] (step3)[base, below of=step2, yshift = -2.5mm]{\textbf{3.} Reject all clusters that contain less than $N_{\rm min}$ points\\or that have median local densities less than $\rho_{\rm threshold}$\\times that at their borders.};
    
    \node[align = center] (step4)[base, below of=step3]{\textbf{4.} Reject all single leaf clusters.};
    
    \node[align = center] (step5)[base, below of=step4, yshift = -2mm]{\textbf{5.} For all parent-child cluster pairs sharing at least\\$f_{\rm reject}$ of the parent's points; reject the child if it\\has child clusters of its own, otherwise reject the parent\\if it has a parent cluster of its own.};
    
    \node[align = center] (step6)[base, below of=step5, yshift = -2.5mm]{\textbf{6.} For all clusters that have a parent cluster or that have\\neither a parent cluster nor any child clusters; reject all\\points, $o$, within them where the $lof(o)>S_{\rm outlier}$.};
    
    \node[align = center] (step7)[base, below of=step6, yshift = -4.5mm]{\textbf{7.} For any cluster whose ordered list's endpoints encompass\\a local maxima whose reachability distance is larger\\than that at these endpoints; reject it if it has child\\clusters, otherwise remove all points from this cluster from\\whichever side of the local maxima contains less.};
    \draw[->](step1) -- (step2);
    \draw[->](step2) -- (step3);
    \draw[->](step3) -- (step4);
    \draw[->](step4) -- (step5);
    \draw[->](step5) -- (step6);
    \draw[->](step6) -- (step7);
\end{tikzpicture}
\end{center}
\vspace{-8pt}
\caption{The cluster extraction activity chart summarising the steps we take to determine clusters from the \code{OPTICS} output. Steps 1, 2, 3, and 6 can be parallelised while steps 4, 5, and 7 can only be partially parallelised -- due to the need to preserve the state of the hierarchy before these steps whilst the step is conducted, refer to step 3 in the main text for more details on this. However, we only perform these steps using a single core due to the entire extraction process only taking a small fraction of the time it takes to run \code{OPTICS} for any particular data set. It is important that these steps be performed exactly in this order else the extracted clusters may not all be meaningful.}
\label{fig:ExtractionProcess}
\end{figure}

Since the \code{OPTICS} algorithm itself does not return any clusters, the extraction of clusters from reachability plots is a separate and unique problem whose difficulties arise due to the innate subjectivity with regards to the definition of a cluster. Two commonly used automatic cluster extraction techniques are the $\xi$-steep method, first proposed in the original \code{OPTICS} paper \citep{Ankerst1999}, and the {\tt DBSCAN} method, which effectively returns those clusters that the \code{OPTICS} predecessor {\tt DBSCAN} \citep{Ester1996} would have. While the $\xi$-steep method is able to extract a hierarchy and the {\tt DBSCAN} method can do so after being applied iteratively, neither is robust enough to extract all necessary clusters at any overdensity. To combat these downfalls, we have developed our own extraction process based on the designs of \citet{Sander2003}, \citet{Zhang2013}, and \citet{McConnachie2018}; Fig. \ref{fig:ExtractionProcess} summarises the steps involved. This extraction process produces a series of tree structures consisting of clusters for nodes which we refer to as the \code{Halo-OPTICS} hierarchy. In alignment with the standard terminology for this data structure, we refer to any pair of clusters separated by a single branch as a parent-child cluster pair, and we refer to a tree's terminating clusters as the root and leaf clusters.

\subsubsection*{\hspace{0.3cm}Step 1}
\vspace{-0.2cm}
\hangindent=0.3cm 
\hspace{0.3cm}Given that dense regions of the data are described by valleys in the reachability plot, our first step to extracting the clusters present in the data is to define all local maxima in the reachability plot as the boundaries of clusters. For the purpose of finding these local maxima, we treat all undefined reachability distances as being equal to $\epsilon$, which occur when a point has never been included in a nearest-neighbour radial query of another point that returns at least $N_{\rm{min}}$ points.

\subsubsection*{\hspace{0.3cm}Step 2}
\vspace{-0.2cm}
\hangindent=0.3cm 
\hspace{0.3cm}We then construct clusters out of lists of consecutively ordered points contained within valleys of the reachability plot. Since each valley must be bordered by a local maximum, we link up contiguous sets of points on both sides of every local maximum in the reachability plot with reachability distance less than that at the corresponding local maximum. This creates two clusters for every local maximum. Following this step, every cluster will contain at least one local minimum, and be bordered by one (possibly two, if a valley is bordered by two local maxima with the same reachability distance) local maximum.

\subsubsection*{\hspace{0.3cm}Step 3}
\vspace{-0.2cm}
\hangindent=0.3cm 
\hspace{0.3cm}As expected, this process incurs many artefact clusters that are insignificant and so we now move through a rejection process that removes these clusters from the list. Since the reachability distance of a particle is the radius of a particle encompassing sphere, we now approximate that the local density is inversely proportional to the volume of the n-sphere -- $n=3$ in this paper -- with a radius of the reachability distance. We use this concept in the first step of our rejection process which rejects any potential clusters that satisfy either of the following: \vspace{0.2cm} \\
(i) The cluster contains less than $N_{\rm{min}}$ particles.\\
(ii) The median local density of the cluster is less than $\rho_{\rm threshold}$ times that at the local maximum that was used to create it. \vspace{0.2cm} \\
The first of these criteria ensures that clusters contain at least $N_{\rm{min}}$ particles as required by the resolution of structures in \code{OPTICS}. For the second criterion, the chosen density contrast, $\rho_{\rm threshold}$, guarantees that half of the points of a cluster must be at least $\rho_{\rm threshold}$ as dense as that cluster's surroundings. Refer to Sec. \ref{subsec:purityandrecovery} for our determination of a reasonable $\rho_{\rm threshold}$ value, ultimately we use $\rho_{\rm threshold} = 2$. Some of the remaining steps are performed by first considering whether each cluster in the list satisfies a condition before then rejecting all such clusters at once. We conduct the process in this way due to some of these conditions depending on the state of the hierarchy and therefore whether or not a cluster satisfies such a condition is susceptible to change under a typical {\it reject mid iteration} type method.

\subsubsection*{\hspace{0.3cm}Step 4}
\vspace{-0.2cm}
\hangindent=0.3cm 
\hspace{0.3cm}We now mark all clusters that are a single child of their parent cluster, before then rejecting each of them. We justify this as a necessary step to remove any clusters that are simply smaller versions of their parent cluster. Single child clusters occur in the hierarchy when one of the two clusters for each local maxima that was originally created in step 2, has been rejected during step 3.

\subsubsection*{\hspace{0.3cm}Step 5}
\vspace{-0.2cm}
\hangindent=0.3cm 
\hspace{0.3cm}At this step, the list of clusters still typically contains many cascading parent-child clusters that share large numbers of points. For all parent-child cluster pairs sharing at least $f_{\rm reject}$ of the parent's points; mark the child for rejection if it has child clusters of its own, otherwise mark the parent for rejection if it has a parent cluster of its own. We then reject those marked clusters after inspecting the entire list. This step further ensures the individuality of clusters in consecutive levels of the hierarchy. Refer to Sec. \ref{subsec:purityandrecovery} for our determination of a reasonable $f_{\rm reject}$ value, ultimately we use $f_{\rm reject} = 90\%$.

\subsubsection*{\hspace{0.3cm}Step 6}
\vspace{-0.2cm}
\hangindent=0.3cm 
\hspace{0.3cm}Another artefact of determining clusters in this way is that each cluster will likely contain outlier points that do not belong as part of the cluster. We now reject outlier particles from all clusters that either; have a parent cluster, or that have neither a parent cluster nor any child clusters. By exempting all root clusters (with child clusters) from the outlier rejection, we maintain the lists of particles that give the best description of larger halo environments. For those clusters that we do apply the outlier detection to, we reject particles on the basis outlined by \citet{Breunig1999} who define the local-reachability-density of a point, $o$, as

\begin{equation} \label{eq:lrd}
\hspace{0.3cm}
    {\rm lrd}(o) = \frac{N_{\rm{min}}}{\sum\limits_{q \in N_{r \leq o.cd}(o)} {\rm reach\_dist}(o, q)}.
\end{equation}

\hangindent=0.3cm 
\hspace{-0.33cm}Here $o.cd = {\rm core\_dist}(o)$ from Eq. \ref{eq:coredistance} and ${\rm reach\_dist}(o, q)$ is defined in Eq. \ref{eq:reachabilitydistance}. It should be noted that these values are found using only the points from within each cluster and in general will be different to those found during the \code{OPTICS} process, and will also differ from cluster to cluster for any point contained in multiple clusters. The local-outlier-factor of $o$ is then defined as

\begin{equation} \label{eq:lof}
\hspace{0.3cm}
    {\rm lof}(o) = \frac{\sum\limits_{q \in N_{r \leq o.cd}(o)} \frac{{\rm lrd}(q)}{{\rm lrd}(o)}}{N_{\rm{min}}}.
\end{equation}

\hangindent=0.3cm 
\hspace{-0.33cm}We find the local-outlier-factor for all points of a cluster, for all clusters. For each cluster, we then reject all points from it that have a local-outlier-factor greater than $S_{\rm outlier}$. Any point that is a part of $n$-many clusters will therefore have $n$ individual local-outlier-factors that are respective to each. It then follows that such a point may be rejected from one cluster and not another. However, it also important to note that since a parent cluster contains a larger number of lower density points than its child cluster, the local-outlier-factor of a point contained in both parent and child clusters will always be larger than or equal with respect to the child cluster than it is with respect to the parent cluster. Therefore following this step, a child cluster will still never contain a point that a parent cluster does not. Refer to Sec. \ref{subsec:purityandrecovery} for our determination of a reasonable $S_{\rm outlier}$ value, although ultimately we use the suggestion from \citet{Breunig1999} that $S_{\rm outlier} = 2$.

\subsubsection*{\hspace{0.3cm}Step 7}
\vspace{-0.2cm}
\hangindent=0.3cm 
\hspace{0.3cm}Following the rejection of outlier points from all clusters, a possible fringe case occurs where the ordered list for some clusters now encompasses one or more points (that are not necessarily contained in the cluster itself) whose local density is less than that of either of the points at the cluster's ordered list bounds. This is essentially a discontinuity in the density field of the cluster as we've determined it thus far. So for any cluster that satisfies this condition, we reject it if it contains a child cluster, otherwise we remove all points from this cluster from whichever side (in the ordered list) of the local maximum contains less of them. If the removal of these points leaves the cluster with less than $N_{\rm{min}}$ particles, then we reject the cluster.\\

\begin{table*}
\centering
\caption{The descriptions and distributional parameters of each of the nine mock clusters used to investigate the purity and recovery statistics of \code{Halo-OPTICS}. The analysis is conducted inside a cube with a side length of $1$ that is centered on the origin. The total number of points within each cluster is $N\times (1-f_b) \times f_c$ rounded to the nearest integer, where $N$ is the total number of points inside the unit cube, $f_b$ is the proportion of background noise, and $f_c$ is the proportion of clustered points belonging to that particular cluster. Notice that $\sum f_c = 1$.}
\begin{tabular}{|c|c|c|c|c|} 
\hline
Cluster & Description & Centre Coordinates & Spread & $f_c$ \\
\hline
A & Sphere & $(x, y, z) = (0, 0, 0)$ & $(\sigma_x, \sigma_y, \sigma_z) = (0.06, 0.06, 0.06)$ & $0.05$ \\
\hline
B & Sphere inside A & $(x, y, z) = (0, 0, 0)$ & $(\sigma_x, \sigma_y, \sigma_z) = (0.01, 0.01, 0.01)$ & $0.25$ \\
\hline
C & Sphere at edge of B & $(x, y, z) = (0.05, 0, 0)$ & $(\sigma_x, \sigma_y, \sigma_z) = (0.005, 0.005, 0.005)$ & $0.25$ \\
\hline
D & Cone extending radially from A & $(x, y, z) = (0.2, 0.2, 0)$ & $(\sigma_r, \sigma_\theta, \sigma_\phi) = (0.05, 2^\circ, 2^\circ)$ & $0.05$ \\
\hline
E & Sphere nearby F & $(x, y, z) = (0.3, 0, 0.03)$ & $(\sigma_x, \sigma_y, \sigma_z) = (0.01, 0.01, 0.01)$ & $0.1$ \\
\hline
F & Sphere nearby E & $(x, y, z) = (0.3, 0, -0.03)$ & $(\sigma_x, \sigma_y, \sigma_z) = (0.01, 0.01, 0.01)$ & $0.1$ \\
\hline
G & Angular arc nearby H & $(x, y, z) = (0, -0.3, 0)$ & $(\sigma_r, \sigma_\theta, \sigma_\phi) = (0.01, 5^\circ, 25^\circ)$ & $0.1$ \\
\hline
H & Sphere inside tail of G & $(r, \theta, \phi) = (0.3, 90^\circ, -135^\circ)$ & $(\sigma_x, \sigma_y, \sigma_z) = (0.01, 0.01, 0.01)$ & $0.02$ \\
\hline
I & Angular arc nearby G & $(x, y, z) = (0, -0.4, 0)$ & $(\sigma_r, \sigma_\theta, \sigma_\phi) = (0.01, 30^\circ, 2^\circ)$ & $0.08$ \\
\hline
\end{tabular}
\label{tab:mockclusters}
\end{table*}

\begin{figure}
    \centering
    \includegraphics[trim={26mm 23mm 12.5mm 14mm}, clip, width=\columnwidth]{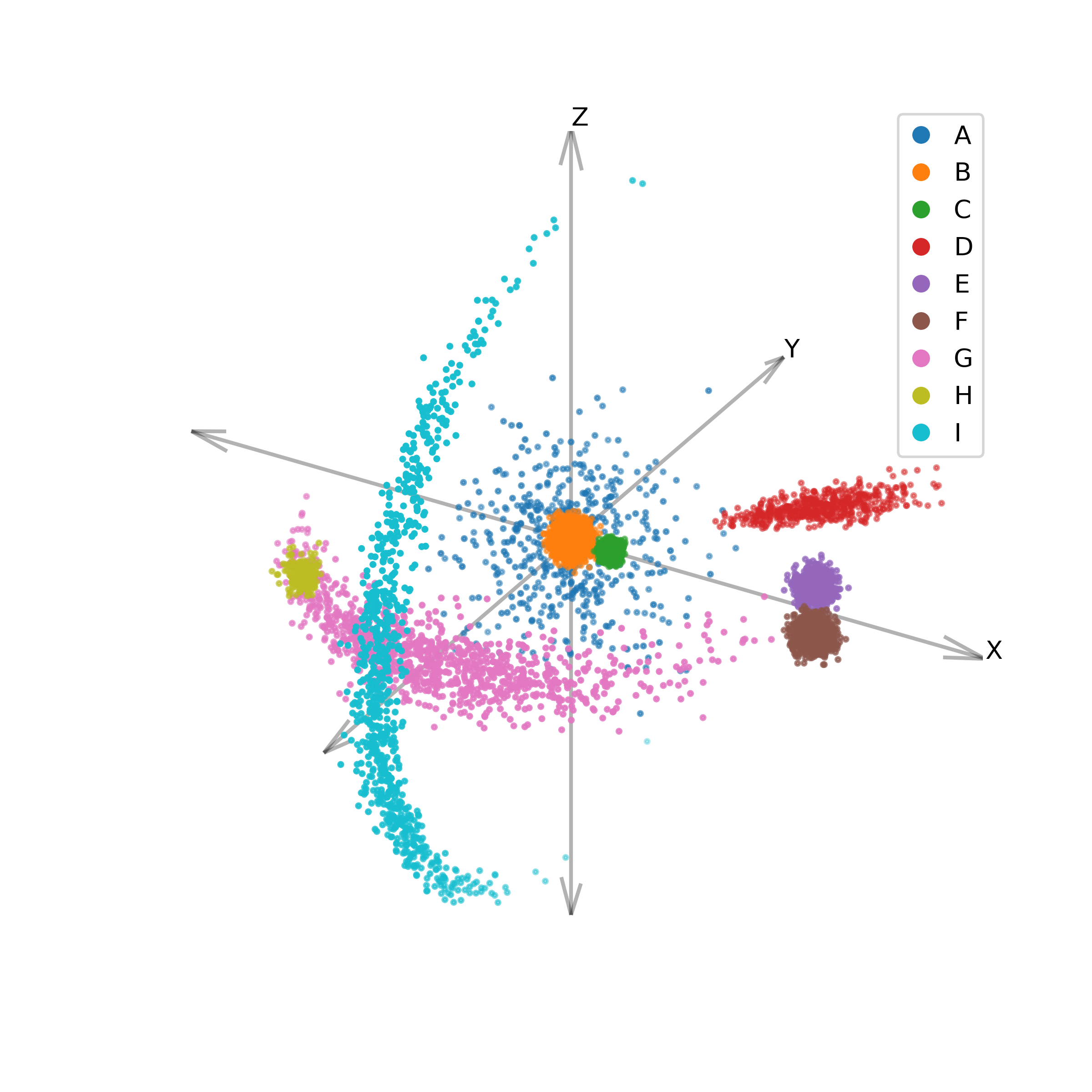}
    \vspace{-0.5cm}
    \caption{Projection of the mock clusters listed in Tab. \ref{tab:mockclusters}. The clusters, which are coloured by their true label, are designed to mimic a variety of typical astrophysical clusters that also provide many intricacies for \code{OPTICS} to interpret such as; closely situated yet unique overdensities, elongated structures, and a (somewhat) arbitrarily multi-levelled hierarchy.}
    \label{fig:mockclusters}
\end{figure}

\begin{figure*}
    \centering
    \includegraphics[trim={12.5mm 5mm 19mm 16.5mm}, clip, width=0.7\textwidth]{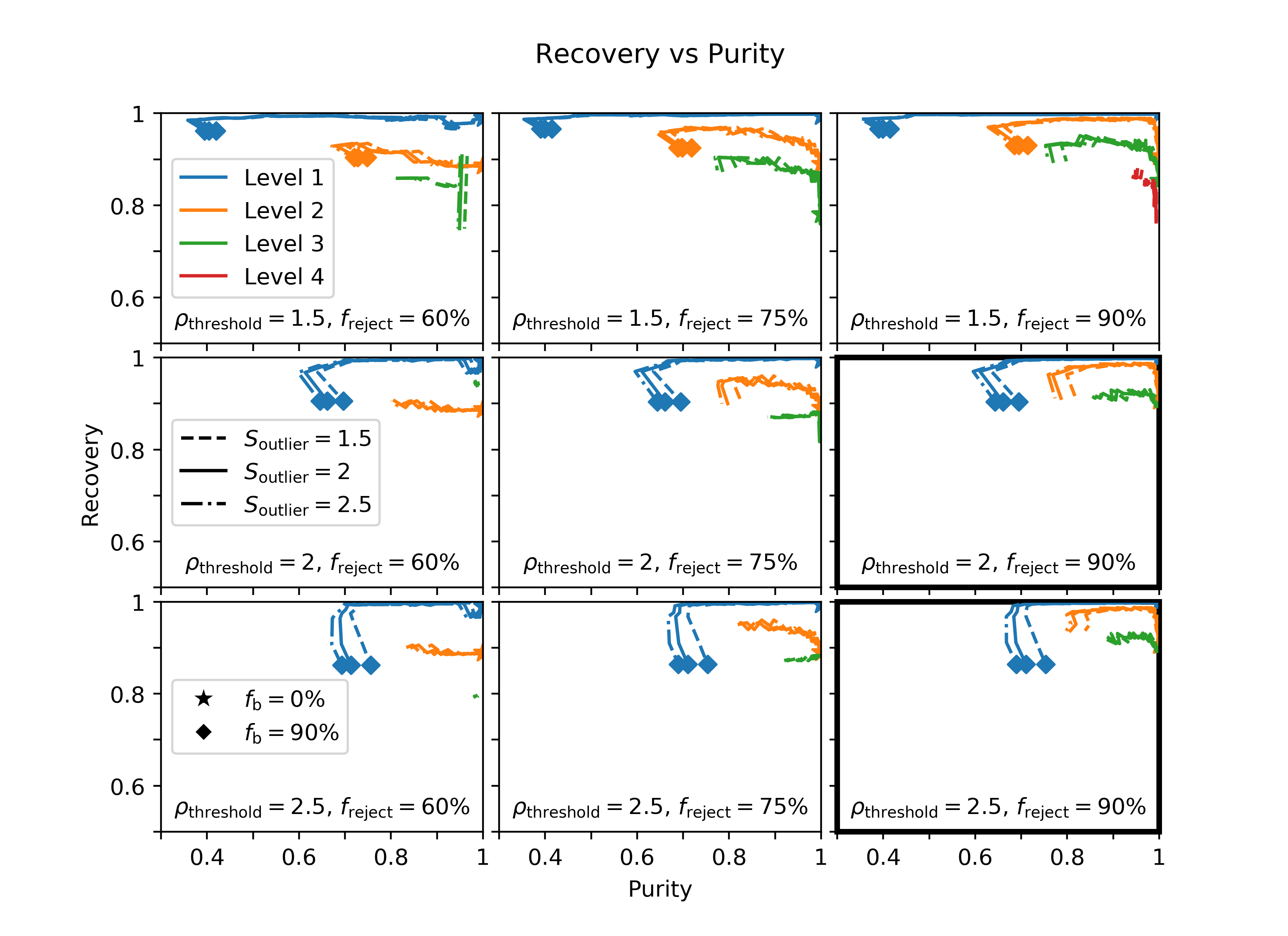}
    \vspace{-0.25cm}
    \caption{The recovery and purity relations of \code{Halo-OPTICS} as dependent on various combinations of the extraction parameters -- $\rho_{\rm threshold}$, $f_{\rm reject}$, and $S_{\rm outlier}$ -- after having been applied to the mock clusters listed in Tab. \ref{tab:mockclusters}. Each panel displays the recovery vs purity for a different $\rho_{\rm threshold}$/$f_{\rm reject}$ combination. Within each panel is the recovery vs purity as it is dependent on the various; $S_{\rm outlier}$ values, hierarchy levels, and levels of background noise. We do not show level 0, the root level, as it contains all points in the data set and hence its recovery vs relation is a trivial line of constant recovery ($=1$) and decreasing purity ($= 1 - f_b/100\%$). We find that the main contribution to the performance of \code{Halo-OPTICS} comes from the $\rho_{\rm threshold}$ and $f_{\rm reject}$ parameters with $S_{\rm outlier}$ making little difference. The bold bordered panels showcase the two best performing parameter combinations which we compare against each other by way of the maximum Jaccard index in Fig. \ref{fig:maxjaccindex}.}
    \label{fig:recoveryvspuritymock}
\end{figure*}

\begin{figure*}
    \centering
    \includegraphics[trim={12mm 11mm 18mm 22.5mm}, clip, width=0.7\textwidth]{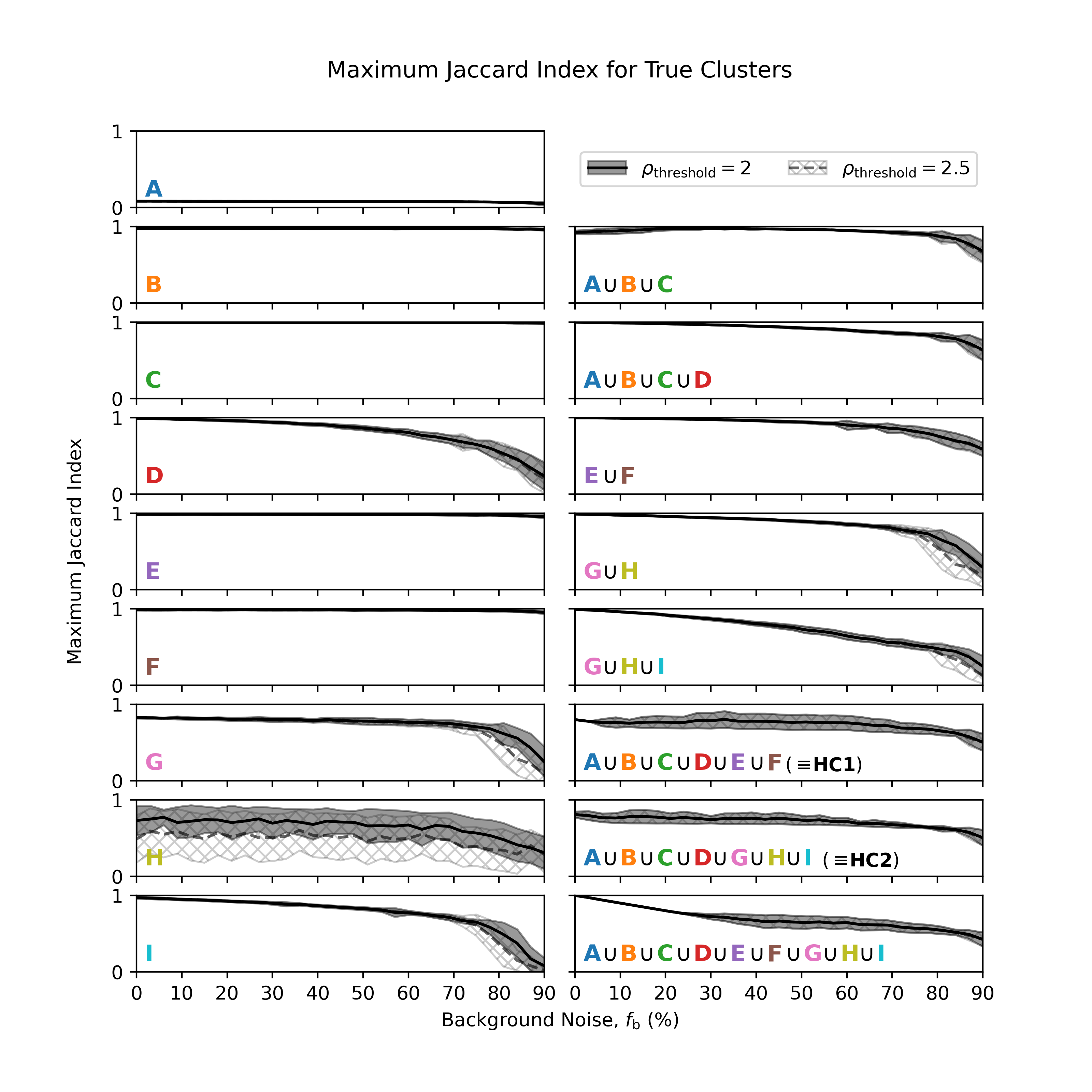}
    \vspace{-0.25cm}
    \caption{The maximum Jaccard index of all mock clusters, and some hierarchical combinations of them, as listed in Tab. \ref{tab:mockclusters}. This measure indicates how close of a match the best-fitting predicted cluster from \code{Halo-OPTICS} is to any particular (union of) true cluster(s). Here we show the mean value and $\mu \pm \sigma$ range of the maximum Jaccard index for both $\rho_{\rm threshold} = 2$ and $\rho_{\rm threshold} = 2.5$ when $f_{\rm reject} = 90\%$ and $S_{\rm outlier} = 2$. Here we see that the denser more spherical clusters are predicted almost perfectly, whereas the more stream-like cluster predictions made by \code{Halo-OPTICS} gradually suffer from the increase in background noise. However, it is obvious from this comparison that $\rho_{\rm threshold} = 2$ does perform better than $\rho_{\rm threshold} = 2.5$ in that the extraction process consistently produces better fitting predictions of the true clusters $D$, $G$, $H$, and $I$ as well as some hierarchical combinations of these. This higher quality performance is particularly noticeable for the more stream-like structures as the background noise increases -- a desired quality of structure finders.}
    \label{fig:maxjaccindex}
\end{figure*}

Following these steps, we are left with a list of clusters that we have determined to be significantly denser, distinct, and self-consistent when compared to their surroundings. Moreover, the process is completed with the addition of only three user defined parameters -- $\rho_{\rm threshold}$, $f_{\rm reject}$, and $S_{\rm outlier}$. It should be noted that the clusters and the hierarchy they are a part of is not necessarily the hierarchy that might be assigned based on physical reasons. It is particularly dependent upon $\epsilon$ in the root level and similarly upon $N_{\rm min}$ in the leaf level. Importantly, the detection of significant overdensities by \code{Halo-OPTICS} is still very informative even though it may not be physical, and further still there is the possibility for implementing changes to the extraction process such that it presents a more physical set of clusters.

\subsection{Performance Optimisation} \label{subsec:purityandrecovery}
To justify our extraction process in Sec. \ref{subsec:structuredetection} we now present some performance statistics of this process and how they are affected by the; extraction parameters $\rho_{\rm threshold}$, $f_{\rm reject}$, and $S_{\rm outlier}$; type of structure present; as well as the level of non-structured background noise within the data set. To do this, we create a mock cluster set designed to mimic typical astrophysical structures. The data are contained within a unit cube centered on the origin and the total number of points is kept at a constant $N = 10^4$. The \code{OPTICS} hyperparameters are chosen as $N_{\rm min} = 20$ to mimic our application to the MW type galaxies and $\epsilon \to \infty$ so that the root cluster includes all points.

The clusters are created using 3D Gaussian distributions of various sizes, spreads and positions in both $x, y, z$ and $r, \theta, \phi$ coordinates\footnote{We do not use typical halo profiles here as the performance of \code{Halo-OPTICS} does not depend on the exact density profile of a cluster. This is due to \code{OPTICS} not using any information about the exact structure of a cluster in order to link the points within it, and is precisely why it excels at finding arbitrarily shaped clusters. We do, however, use a typical halo profile for clusters in Sec. \ref{subsec:hierarchyexplanation} as the exact density profile will affect the conditions under which particle bridges are created -- which in turn affects the shape of the \code{Halo-OPTICS} hierarchy.}. The descriptions and distributional parameters of these clusters are presented in Tab. \ref{tab:mockclusters}, and a 2D projection of one sampling is shown in Fig. \ref{fig:mockclusters}. The proportion of the total clustered points is given by $(100\% - f_b)$ where $f_b$ is the percentage of background noise. We vary the background noise from $f_b = 0\%$ to $f_b = 90\%$ in increments of $3\%$ and sample it using a uniform distribution of points throughout the space. For each level of $f_b$ we run \code{OPTICS} $50$ times, re-sampling from all distributions for each run.

We first assess the performance of \code{Halo-OPTICS} through measures of recovery and purity. We define the recovery to be a function of the true clusters and to be dependent on the levels of the predicted hierarchy such that
\begin{equation} \label{eq:recovery}
    {\rm R}(T|L) = \frac{\sum\{|T\cap C| \mid \forall C \in L\}}{|T|}, T \in M \land L \subset H.
\end{equation}
Here $T$ is a true (mock) cluster, $M$ is the set of mock clusters, $C$ is a \code{Halo-OPTICS} predicted cluster, and $L$ is the set of predicted clusters that belong to the $L^{\rm th}$ level of the predicted hierarchy, $H$. This way the recovery of a particular true cluster can be interpreted as the fraction of that cluster that is returned in the $L^{\rm th}$ level of the predicted hierarchy. Since predicted clusters from \code{Halo-OPTICS} in the same hierarchical level cannot overlap, this value will always be contained to the interval $[0, 1]$.

Similarly, we define the purity to be a function of the predicted clusters and to be dependent on the levels of the predicted hierarchy such that
\begin{equation} \label{eq:purity}
    {\rm P}(C|L) = \frac{\sum\{|T\cap C| \mid \forall T \in M\}}{|C|}, C \in L \subset H.
\end{equation}
Here $T$, $M$, $C$, $L$, and $H$ are the same terms as in Eq. \ref{eq:recovery}. This definition of the purity of a particular predicted cluster, $C$, can be interpreted as the fraction of $C$ that is not background noise.

For each of the background noise/distribution re-sampling combinations we find the recovery vs purity relations for $27$ \code{Halo-OPTICS} parameter combinations. The combinations draw from $\rho_{\rm threshold} \in \{1.5, 2, 2.5\}$, $f_{\rm reject} \in \{60\%, 75\%, 90\%\}$, and $S_{\rm outlier} \in \{1.5, 2, 2.5\}$. Whilst masking all zero recovery and zero purity values, we then average over both the distributional re-samplings and each level of the predicted hierarchy.

Fig. \ref{fig:recoveryvspuritymock} depicts these recovery vs purity relations as dependent on the parameter combination, the level of the hierarchy, and the level of background noise. For all parameter combinations the recovery and purity decrease and increase respectively for a given level of background noise as the level of the hierarchy deepens. This should be expected, clusters in deeper levels of the hierarchy are denser and have fewer points than their parent clusters. As a result, they are less affected by the increase in noise (higher purity per level of background noise) and are less likely to include more of the total clustered points in the data set (lower recovery in general).

Another noticeable feature in Fig. \ref{fig:recoveryvspuritymock} is that some deep levels of the hierarchy reveal a drop in recovery at $f_b = 0\%$. This characteristic is exaggerated for low $\rho_{\rm threshold}$ and low $f_{\rm reject}$ values. Following step 1 of the extraction process, the valley of the reachability plot that corresponds with any particular true cluster will typically have some non-zero number of cascading parent-child predicted clusters associated with it -- these range from high density with a few points, to lower density with more points. In general, lowering $\rho_{\rm threshold}$ allows for the extraction of more higher density child clusters within this valley, and lowering $f_{\rm reject}$ effectively removes the parent clusters above them. Increasing the level of background noise lowers the density contrast surrounding the each of the mock clusters, which in turn stops the low $\rho_{\rm threshold}$ from allowing the extraction of as many child clusters. Likewise, the resulting predicted cluster has a higher recovery and slightly lower purity for some $f_b > 0\%$ since it contains more points at a lower density. Eventually, as the background noise level increases, the prediction of the true clusters breaks down and the recovery drops dramatically. This can be seen at each hierarchy level for every parameter combination.

We see here that the choice of $S_{\rm outlier}$ makes little difference to the recovery vs purity relation and as such we choose to take $S_{\rm outlier} = 2$, the suggestion by \citet{Breunig1999}. Effectively, $S_{\rm outlier}$ is only responsible for removing a small number of points in comparison the size of the cluster, so it should be expected that this parameter has little effect on the recovery and purity. The bold bordered panels are the best performing $\rho_{\rm threshold}$/$f_{\rm reject}$ parameter combinations. It is also clear here that $f_{\rm reject} = 90\%$ is a good choice, however it is not immediately obvious as to whether $\rho_{\rm threshold} = 2$ or $\rho_{\rm threshold} = 2.5$ is a better choice.

To help distinguish between the two $\rho_{\rm threshold}$ choices, we also look at the maximum Jaccard index for each true cluster as well as some hierarchical combinations of them. The maximum Jaccard index is defined as
\begin{equation} \label{eq:maxjaccindex}
    {\rm J}_{\rm max}(T) = {\rm max}\bigg\{ \frac{|T\cap C|}{|T\cup C|} \mid \forall C \in H \bigg\}, T \in M^\prime.
\end{equation}
Here $M^\prime$ is the set of true clusters as well as some typically predicted hierarchical combinations of them. Likewise, $T$ is any element of $M^\prime$, and $C$ is any cluster in the predicted hierarchy, $H$, regardless of the hierarchy level it belongs to. The maximum Jaccard index provides a measure of {\it how close of a match does the best-fitting predicted cluster provide for any given true cluster} -- a true cluster here being any element of $M^\prime$. In this way, we can test the performance of \code{Halo-OPTICS} by examining how well it predicts $M^\prime$. Fig \ref{fig:maxjaccindex} depicts the mean and one standard deviation range of each $T \in M^\prime$ for both $\rho_{\rm threshold} = 2$ and $\rho_{\rm threshold} = 2.5$ when $f_{\rm reject} = 90\%$ and $S_{\rm outlier} = 2$. Here the mean and standard deviation are found from the series of \code{Halo-OPTICS} outputs over each of the distributional re-samplings.

These also reveal a few features of the \code{Halo-OPTICS} predicted clusters in general. One such feature is that the stream-like structures -- $D$, $G$, and $I$ -- are more affected by the increase in background noise. This could be expected since these structures are elongated and will have a lower spatial density than a Gaussian sphere, and therefore will become less distinguishable from the background noise more readily. Another feature is over-encompassing clusters, such as cluster $A$, are not shown to be well-matched here and have a lower maximum Jaccard index than other clusters. This is due to the algorithm not separating out the inner clusters, $B$ and $C$, and in this way the best-fitting match for cluster $A$ is most probably that which provides the best match to the hierarchical combination of clusters $A$, $B$, and $C$. Of course, if the set difference of the best-fitting match to the latter was taken with the best-fitting matches to clusters $B$ and $C$, we could provide a better match to cluster $A$ alone. Interestingly, the maximum Jaccard index of the hierarchical combination of clusters $A$, $B$, and $C$ has a slight peak around $f_b \approx 30\%$. This occurs as a result of the reachability distance for the outer points of cluster $A$ being reduced by the addition of the background noise points in these regions. This effect is small and, as the background noise increases, is outweighed by the inclusion of additional intra-cluster noise.

Fig. \ref{fig:maxjaccindex} also includes two panels of a pair of disjoint hierarchical combinations of clusters -- namely the union of clusters $A$, $B$, $C$, $D$, $E$, and $F$, and the union of clusters $A$, $B$, $C$, $D$, $G$, $H$, and $I$. These have been included to express that the purposely arbitrary hierarchy we have constructed within our mock cluster set has been translated into an equally ambiguous predicted hierarchy. For the purpose of providing an explanation with regards to this, we will refer to the former hierarchical combination as HC1 and the latter as HC2 for the remainder of this paragraph. The disjoint nature of HC1 and HC2 ensures that their respective best-fitting predicted clusters cannot be one and the same unless their shared best-fitting predicted cluster either; contains only those leaf clusters that are common to both HC1 and HC2 (i.e. $A$, $B$, $C$, and $D$), or contains all leaf clusters (i.e. $A$ through to $I$). In the following we refer to these scenarios as case 1 and case 2, respectively\footnote{There are technically more cases that can occur, for example, where HC1 (or HC2) is best-matched by a predicted cluster that includes only points from the leaf clusters within it (and some level of background noise), and then HC2 (or HC1) is best-matched by either of the predicted clusters in cases 1 and 2. While this occurs for some other hyperparameter combinations that return a larger number of hierarchy levels, it does not occur within the hyperparameter combinations featured in Fig. \ref{fig:maxjaccindex} -- and so we do not discuss these extra cases.}. For $f_b = 0\%$, the Jaccard index for cases 1 and 2 is $0.6/0.8 = 0.75$ and $0.8/1 = 0.8$ respectively, and hence the maximum Jaccard index for both HC1 and HC2 at this level of background noise is $0.8$. However, as the background noise level increases the Jaccard index for each of these cases changes in a way that depends on the effective occupied volume with the unit cube of the true and predicted clusters proportionally -- there is also some random component to this due to the randomised sampling of both the true clusters and the background noise. The occupied volume of the predicted clusters changes with the background noise as well. As a result, the best-fitting predicted cluster is not always provided by that from case 2, and the corresponding maximum Jaccard index to the best-fitting predicted cluster is not confined to less than $0.8$ either. This is shown by the increase in the spread of the maximum Jaccard index for both HC1 and HC2.

From Fig. \ref{fig:maxjaccindex}, we clearly see that with $\rho_{\rm threshold} = 2$, \code{Halo-OPTICS} out performs that with $\rho_{\rm threshold} = 2.5$ for true clusters $D$, $G$, $H$, and $I$ -- as well as some hierarchical combinations of the latter three -- particularly when the background noise dominates the data. Not only is the mean value of the maximum Jaccard index for these clusters larger under the $\rho_{\rm threshold} = 2$ parameter scheme, but the spread is smaller too. It is now apparent that $\rho_{\rm threshold} = 2$ is the better parameter choice. So to summarise, we use $\rho_{\rm threshold} = 2$, $f_{\rm reject} = 90\%$, and $S_{\rm outlier} = 2$ as our (near) optimal \code{Halo-OPTICS} hyperparameters.

\begin{table}
\centering
\caption{Details of the number of clusters, $|H|$, in the \code{Halo-OPTICS} hierarchy and the dependency of this on the resolution, mass fraction, and separation distance variables of the two-profile system described in Sec. \ref{subsec:hierarchyexplanation}. Due to the non-linear dependencies between the $4$, discretely sampled, variables we only give a qualitative description of the generalised domain for which a particular number of clusters may be found in the input space. For the purposes of being succinct, we refer to the resolution, mass fraction, and separation distance as $R$, $M$, and $S$ respectively.}
\begin{tabular}{|c|c|c|} 
\hline
$|H|$ & Occurrences & Domain Description \\
\hline
$0$ & $417$ & Very small $R$ $\land$ very large $M\ \land$ large $S$ \\
\hline
$1$ & $10471$ & $($Small $R$ $\land$ small $M)\ \lor\ ($very small $S)$ \\
\hline
$2$ & $484$ & Mid-range $R$ $\land$ mid-range $M\ \land$ large $S$ \\
\hline
$3$ & $2488$ & Mid-range $R$ $\land$ mid-range $M\ \land$ mid-range $S$ \\
\hline
$4$ & $209$ & Very large $R\ \land$ very large $M\ \land$ large $S$ \\
\hline
$5$ & $374$ & Very large $R\ \land$ very large $M\ \land$ mid-range $S$ \\
\hline
\end{tabular}
\label{tab:hierarchydependencies}
\end{table}

\subsection{Understanding the \textsc{Halo-OPTICS} Hierarchy} \label{subsec:hierarchyexplanation}

We now wish to inform the reader about the nature of the \code{Halo-OPTICS} hierarchy. To do this we construct another mock example of clusters, only now we intend for these clusters to constitute an easily understandable hierarchy. The mock example we use here consists of two distributions that are intended to represent a main halo and a satellite halo. Both of these haloes are modelled using the spherical NFW profile \citep{Navarro1996} which has a density profile of the form
\begin{equation} \label{eq:nfwprofile}
    \rho(r) = \frac{\rho_0}{\frac{r}{R_s}\left(1 + \frac{r}{R_s}\right)^2},
\end{equation}
where $\rho_0$ and $R_s$ are the characteristic density and radius respectively. We use this to create a cumulative distribution function for the NFW profile from which to sample the main and satellite halo distributions from. We integrate the mass density profile in Eq. \ref{eq:nfwprofile} over the volume and out to some variable radius $r$, and then re-normalise this such that the integral out to some maximum radius, $R_{\rm max}$, is unity. It then follows that an appropriate cumulative distribution function for the haloes is given by
\begin{equation} \label{eq:nfwcdf}
    F_{\rm NFW}(r) = \frac{\left[ln\left(1 + \frac{r}{R_s}\right) - \frac{r}{(r + R_s)}\right]}{\left[ln\left(1 + \frac{R_{\rm max}}{R_s}\right) - \frac{R_{\rm max}}{(R_{\rm max} + R_s)}\right]},
\end{equation}
where $0 \leq r \leq R_{\rm max}$. For the main halo we choose $R_s = 1$ and for the satellite halo we choose $R_s = 0.2$. For both haloes we use $R_{\rm max} = 10R_s$ and for the main halo we treat the effective $R_\Delta$ as being equal to $5R_s$ for the purposes of constructing $\epsilon$ from the data in the way described in Sec. \ref{subsec:opticsparams}.

\begin{figure*}
    \centering
    \includegraphics[trim={12mm 3mm 17mm 14mm}, clip, width=0.7\textwidth]{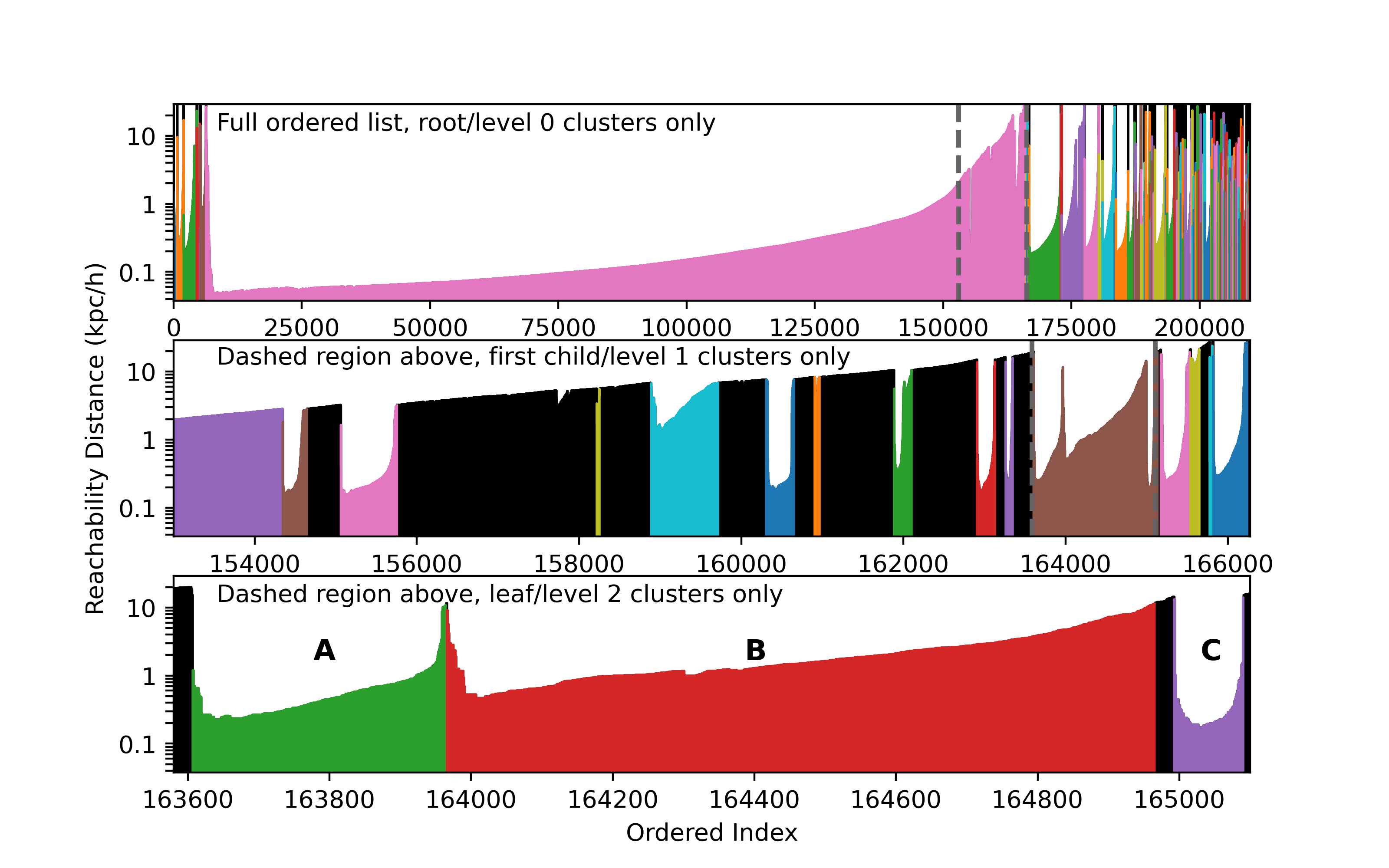}
    \vspace{-0.15cm}
    \caption{The reachability plot of the stellar particles from the MW02 synthetic halo where the colours indicate different clusters as determined by the extraction process outlined in Sec. \ref{subsec:structuredetection}. The top panel is the full reachability plot and has only the root clusters coloured. The middle panel is the section of the reachability plot that is marked with the grey dashed lines in the top panel. The clusters in this panel are coloured at the first child level below the root level. The bottom panel is the section of the reachability plot that is marked with the grey dashed lines in the middle panel. The clusters in this panel are coloured at the second child level below the root level. The black regions within each panel correspond to unclustered points at that level of the hierarchy. The bold letter labels in this bottom panel correspond to the reachability profile of those clusters that have been similarly marked in Fig. \ref{fig:RDClusters}.}
    \label{fig:ReachabilityPlot}
\end{figure*}

\begin{figure*}
    \centering
    \includegraphics[trim={13cm 17cm 18.5cm 13cm}, clip, width=0.7\textwidth]{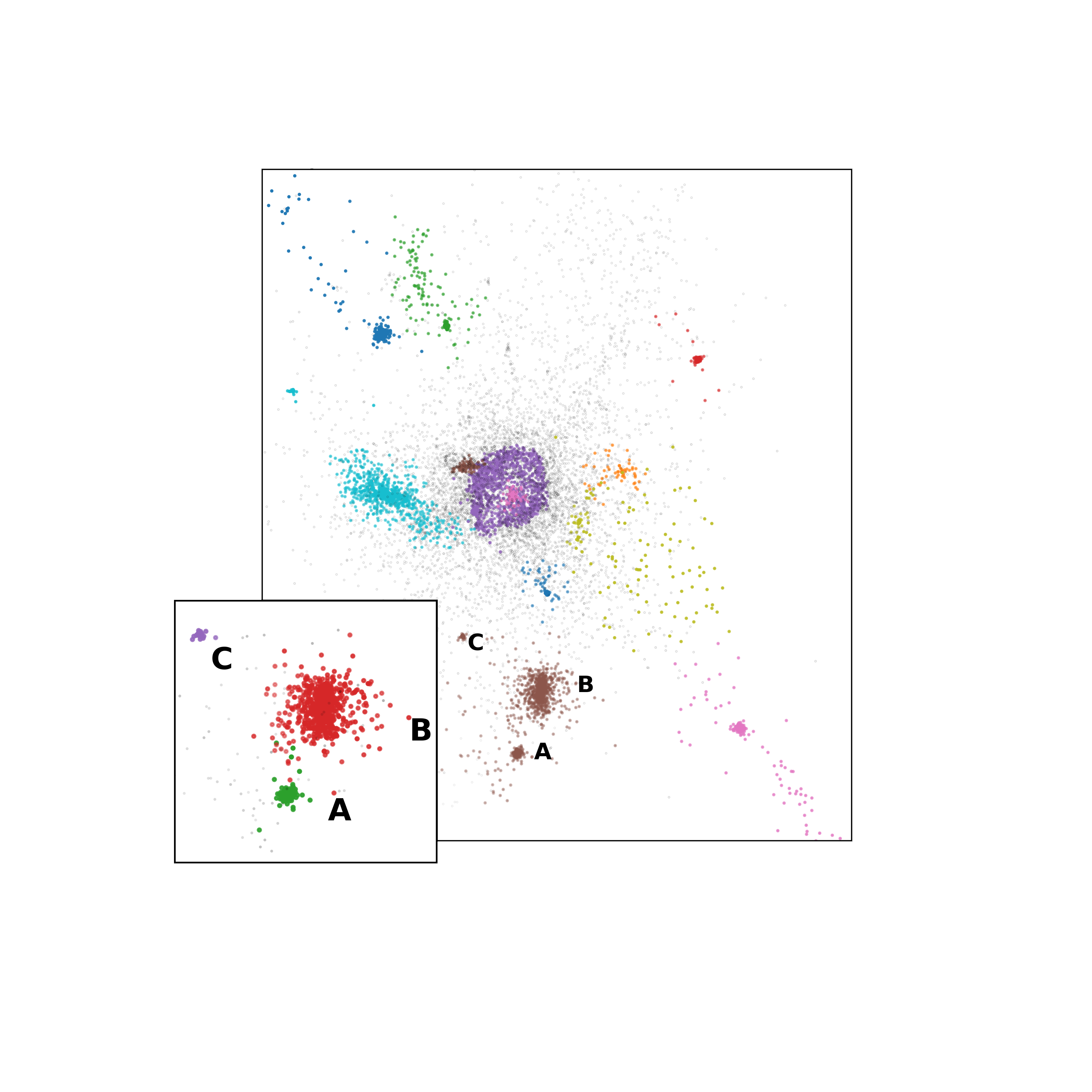}
    \vspace{-0.15cm}
    \caption{The main and inset panels are the positions -- with the corresponding colour schemes -- of all points in the middle and lower panels of Fig. \ref{fig:ReachabilityPlot} respectively. Here we showcase that \code{Halo-OPTICS} retrieves the relevant clusters as well as the appropriate hierarchy that they are contained within.}
    \label{fig:RDClusters}
\end{figure*}

We now vary the resolution, mass fraction, and separation distance of the two-halo system and run \code{Halo-OPTICS} over each realisation using the previously mentioned values of the \code{Halo-OPTICS} hyperparameters. The resolution is the total number of points in the data set which we sample from $13$ logarithmically spaced values between $40$ and $10^4$. The mass fraction is the ratio between the number of points in the satellite and the number of points in the main halo which we sample from $11$ logarithmically spaced values between $0.002$ and $1$. These values correspond to the number of particles belonging to the satellite being equal to; the lowest possible limit of detection by \code{Halo-OPTICS} when the resolution is $10^4$ (when using $N_{\rm min} = 20$); equal to the number of particles in the main halo. The separation distance is the distance between the centres of the two distributions which we sample from $101$ linearly spaced values from $0$ to $20$ -- so as to provide a reasonable range of possible hierarchies.

From Tab. \ref{tab:hierarchydependencies} we see the dependencies of the size of the \code{Halo-OPTICS} hierarchy, $|H|$, upon the input space defined above. The exact number of input space combinations that correspond to a particular value of $|H|$ is not as important as the domains within which these particular values of $|H|$ occur. The case where $|H| = 0$ reflects the scenario where \code{Halo-OPTICS} is unable to gather any grouping of $N_{\rm min}$ points within a radius of $\epsilon$. As such, this occurs at a small resolution, large mass fraction, and large separation distance -- which effectively spreads a small number of points among two distinctly separate and equally massive distributions. The case of $|H| = 1$ typically occurs due to either, or a combination, of a small resolution and a small mass fraction. The fact that it is the most commonly occurring case is simply an artefact of having performed a logarithmic sampling of both the resolution and mass fraction variables -- which has artificially created a sampling bias towards the smaller values of these variables. A single cluster may also be returned for very small separation distances. As such, these variable ranges force \code{Halo-OPTICS} to ignore the satellite in the system and only find the points from the main halo to be significantly clustered. This is a consequence of the satellite having too few points associated with it and/or the two density profiles being indistinctly separated from each other.

The case where $|H| = 2$ is simply a partitioning of the main and the satellite halo distributions into $2$ root clusters and hence occurs at large separation distances -- provided that the resolution and mass fraction variables are large enough to create significant samplings of both the main and satellite halo distributions. The case where $|H| = 3$ generally occurs for the mid-range values of the separation distance -- again, provided that the resolution and mass fraction variable values are suitable. A hierarchy consisting of $3$ clusters marks the case where the main and satellite halo distributions are connected via the means of a particle bridge, whilst still having the two density peaks remain distinctly separate. In this scenario, \code{Halo-OPTICS} finds both the sampled distributions to be leaf clusters of an overarching root cluster. In a real astrophysical data set, such a root cluster would have a particular overdensity that is related to the mapping from $\epsilon$ to the overdensity factor $\Delta$ in the way that is outlined in Sec. \ref{subsec:opticsparams}.

The mock system we investigate here contains $2$ NFW profiles which can only be hierarchically connected via a single overarching root cluster, and as such our explanations thus far have covered all hierarchies that should be expected from the density profile of this system. However, since we perform a random and discrete sampling of these distributions in order to construct the system, the hierarchies that are feasibly possible here extend out to larger sizes than this -- although they are strongly probabilistic in nature. All occurrences of hierarchies in which more than $3$ clusters are found are due to the presence of randomly occurring clusters (ROCs) and are an artefact of the combination of the largest values of both the resolution and mass fraction variables. Together these effectively increase the probability that at least $N_{\rm min}$ points will form a randomly positioned and distinct overdensity within the system. When such a ROC is found, it is forced to a deeper level of the hierarchy as a leaf cluster, alongside another leaf cluster which conforms to the remainder of the sampled distribution that occupies the region that is denser than the saddle point of the density field -- this saddle point is produced due to the presence of the ROC.

It is apparent that the general domain of these larger hierarchies becomes increasingly restricted towards the largest values of the resolution and mass fraction variables as $|H|$ increases. The general decrease in the occurrences for larger hierarchies is due to the probability of a ROC being found within this input space. As such, the cases where $|H| = 4$ and where $|H| = 5$ are effectively extensions of the $|H| = 2$ and $|H| = 3$ cases into this domain respectively -- i.e. the same separation distance ranges with the larger resolution/mass fraction values. Even larger values of $|H|$ are also possible within this input space, although the probability of their occurrence is much lower. It should be noted that in this mock system, we could decrease the likelihood of the ROCs being found by simply increasing the \code{Halo-OPTICS} hyperparameter, $\rho_{\rm threshold}$. However, the value of $\rho_{\rm threshold} = 2$ has a particular significance when pertaining to the detection and extraction of streams. Furthermore, the legitimacy of ROCs changes and becomes somewhat ambiguous when a true astrophysical data set is concerned. Nevertheless, this result may hint at the benefit that \code{Halo-OPTICS} stands to gain from some extra hierarchical cleaning processes.

\section{Output Analysis} \label{sec:results}
\subsection{Visualising the \textsc{Halo-OPTICS} Output} \label{subsec:opticsoutput}
The 3D positions and masses for the stellar and dark matter particles are taken from the MW02, MW03, MW04, and MW06 synthetic haloes simulated by \citet{Power2016}. We use \code{Halo-OPTICS} on the stellar and dark matter particle types within each halo both separately and combined. \code{Halo-OPTICS} gives us a reachability plot (detailed in Sec. \ref{subsec:optics}) as well as a list of significant clusters extracted from that (detailed in Sec. \ref{subsec:structuredetection}). The reachability plot for the stellar particles from the MW02 halo is shown in Fig. \ref{fig:ReachabilityPlot}. The three panels display various ranges of the ordered index of particles, and the colours within each panel are cyclic between consecutive significant clusters of the root, first child, and second child levels from top to bottom respectively. Fig. \ref{fig:RDClusters} contains a main panel and an inset one that depict the positions of the particles contained within the middle panel and bottom panels of Fig. \ref{fig:ReachabilityPlot} respectively. The colour scheme and bold letter labels within these two panels correspond to those used in the middle and bottom panels of Fig. \ref{fig:ReachabilityPlot} respectively.

It can be seen in Fig. \ref{fig:ReachabilityPlot} that even though the reachability plot is not smooth, our extraction process retrieves the more significant clusters while ignoring smaller undulations and noise. The reachability plot from each of the stellar and dark matter particle runs for each of the synthetic haloes always contain a very large valley relating to the denser region that surrounds the MW type galaxy. Within this root level valley there is another very large valley -- as well as many other smaller ones. This {\it other} very large valley is the inner halo (the right edge of which can be seen on the left of the middle panel in \ref{fig:ReachabilityPlot}), which is the most massive leaf cluster present within each synthetic halo. The reachability distance of the inner halo is not only small, indicating a high density, but also quite smooth. The central region of Fig. \ref{fig:RDClusters} features the MW02 inner stellar halo's least dense particles (i.e. those that possess the largest reachability distances). Towards the most dense parts of this inner halo, there is typically many small sharp peaks in reachability (not visible at the scale of the top panel in Fig. \ref{fig:ReachabilityPlot}) however these are artefacts of the \code{OPTICS} process and occur due to the reachability distance not being updated for points that have already been appended to the ordered list. Our extraction process successfully ignores these artefacts.

\begin{figure*}
    \centering
    \includegraphics[trim={10.5mm 0mm 7.5mm 9mm}, clip, width=0.75\textwidth]{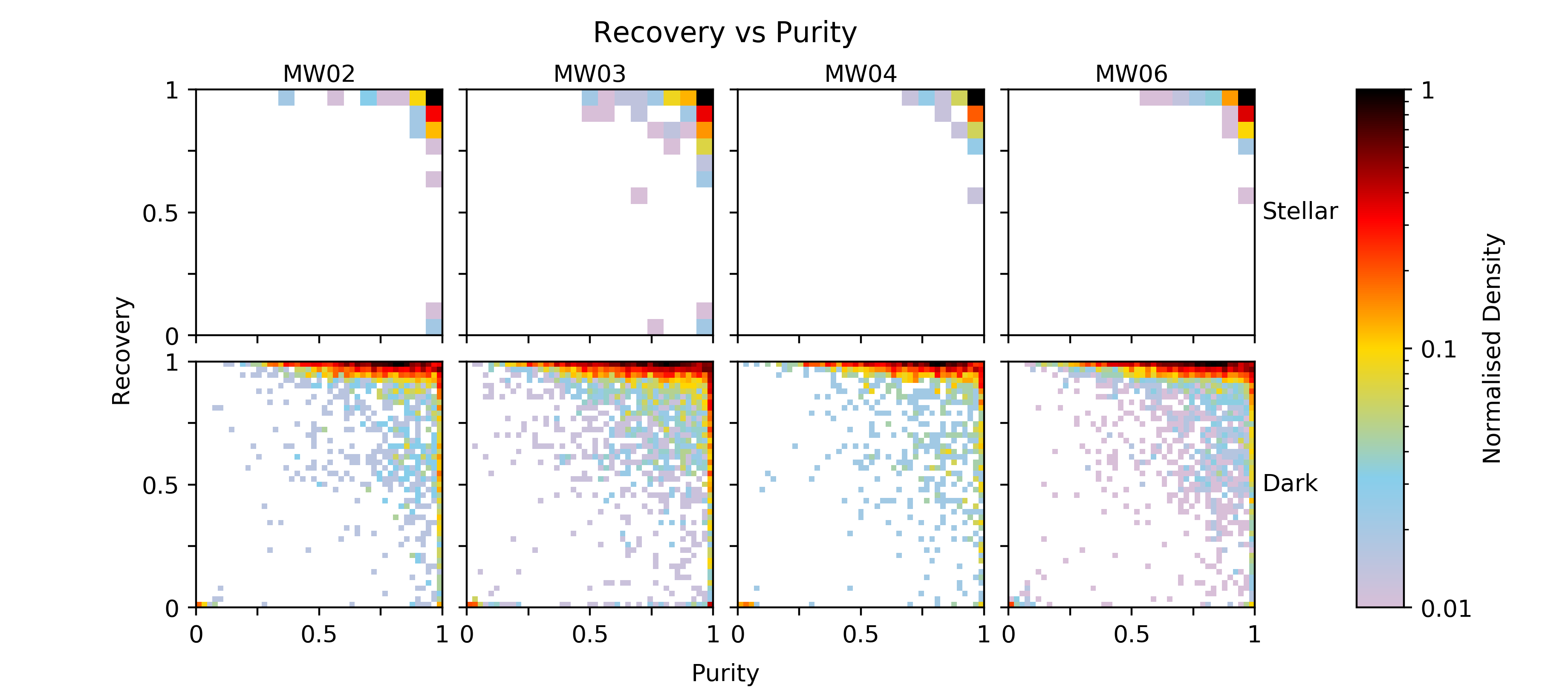}
    \vspace{-0.25cm}
    \caption{The recovery vs purity relations for all \code{Halo-OPTICS} clusters and their best-fitting (by means of the maximum Jaccard index) \code{VELOCIraptor} clusters. Each column displays the relations for a different MW galaxy analogue and the rows correspond to the two particle types -- stars and dark matter. The colour within each panel reflects the normalised density (such that the maximum is $1$) of best-fitting cluster pairs within the cells of a $15$ by $15$ histogram for the stellar clusters and within a $45$ by $45$ histogram for the dark matter clusters.} Within each panel the; upper right; lower right; upper left; and lower left corners indicate the regions that a \code{Halo-OPTICS} cluster will be placed if it is; well-matched by; contained within and comparatively smaller than; containing and is comparatively larger than; mostly (or completely) unrecovered by; its best-fitting \code{VELOCIraptor} cluster. It is shown that a large portion of the clusters from \code{Halo-OPTICS} are well matched by the \code{VELOCIraptor} catalogue with high recovery and purity. We also see here that the \code{Halo-OPTICS} dark matter clusters do have a more variable purity than that of their stellar counterparts when comparing to the \code{VELOCIraptor} output, however these clusters do mostly have high recovery and high Jaccard indices.
    \label{fig:recoveryvspurityvr}
\end{figure*}

\subsection{A Comparison with \textsc{VELOCIraptor}} \label{subsec:comparison}
We now wish to inform the reader as to what the structures are that \code{Halo-OPTICS} has found. We do this via a comparison with the state-of-the-art galaxy/(sub)halo finder \code{VELOCIraptor}. We apply \code{VELOCIraptor} to both the stellar and dark matter particle types within each MW type galaxy's snapshot file. \code{VELOCIraptor} first searches for field haloes from particle positions using a 3D \code{FOF} algorithm. Then, each field halo is searched for phase-space overdensities using an adaptive 6D \code{FOF} algorithm where the position and velocity linking lengths are based on the average spatial and kinematic dispersions of the parent cluster. These linking lengths are iteratively decreased in order to identify local maxima in phase-space density (cores) until no local maxima with enough particles are found. Particles of the root cluster are then iteratively assigned to their nearest core in phase-space, according to the core's phase-space dispersion tensor. A core's phase-space dispersion tensor is updated as new particles are assigned. This process is similar to a Gaussian mixture model but where the number of distributions is fixed to the number of significant phase-space cores found.

To appropriately compare the two codes, we find the best-fitting match from the \code{VELOCIraptor} catalogue of clusters for each \code{Halo-OPTICS} cluster, which we do by means of the maximum Jaccard index, described in Eq. \ref{eq:maxjaccindex}. Then for each \code{Halo-OPTICS}-\code{VELOCIraptor} best-fitting pair we compute the recovery and purity fractions. Fig. \ref{fig:recoveryvspurityvr} depicts the recovery vs purity fractions for all \code{Halo-OPTICS} clusters as compared to their best-fitting \code{VELOCIraptor} cluster for both stellar and dark matter particle types within each MW type galaxy's snapshot file.

We see from these that for a large majority of \code{Halo-OPTICS} clusters there is a good match present within the corresponding \code{VELOCIraptor} catalogue. Of the \code{Halo-OPTICS} clusters that have not been well-matched by \code{VELOCIraptor}, there is a small portion that have high purity and low recovery. We note that of these, many sit deep within the \code{Halo-OPTICS} hierarchy and have a best-fitting \code{VELOCIraptor} cluster that sits comparatively towards the root clusters of the \code{VELOCIraptor} hierarchy. The particles within these clusters are likely to be spatially clustered and not kinematically clustered, hence \code{VELOCIraptor} does not find them to be significantly clustered at a deeper level of its hierarchy. In this scenario it is probable that, given the same particle information as \code{VELOCIraptor}, \code{Halo-OPTICS} would too find these clusters to be insignificant.

It is particularly striking to see that \code{Halo-OPTICS} does quite well at retrieving a large majority of \code{VELOCIraptor} clusters -- an impressive result considering \code{VELOCIraptor}'s use of particle kinematics in order to find many of these. These clusters must still have a significant spatial density contrast with respect to their background to be detected by \code{Halo-OPTICS} but some fine tuning would be needed in order to retrieve these clusters with just a 3D \code{FOF} algorithm. This hints at the greater detection and extraction power of the comparatively adaptive \code{Halo-OPTICS} algorithm over that of a static \code{FOF} algorithm. \code{VELOCIraptor} overcomes this problem by searching through a data-driven position-velocity phase-space to further separate these clusters from their background -- which does not require as much fine-tuning to achieve. Since \code{Halo-OPTICS} does not require this kind of fine-tuning, these results bode well for the performance of \code{Halo-OPTICS} in the event that it is applied using a more informative metric -- inclusive of particle kinematics and metallicities for example.

\begin{figure*}
    \centering
    \includegraphics[trim={12mm 3mm 20mm 14mm}, clip, width=0.7\textwidth]{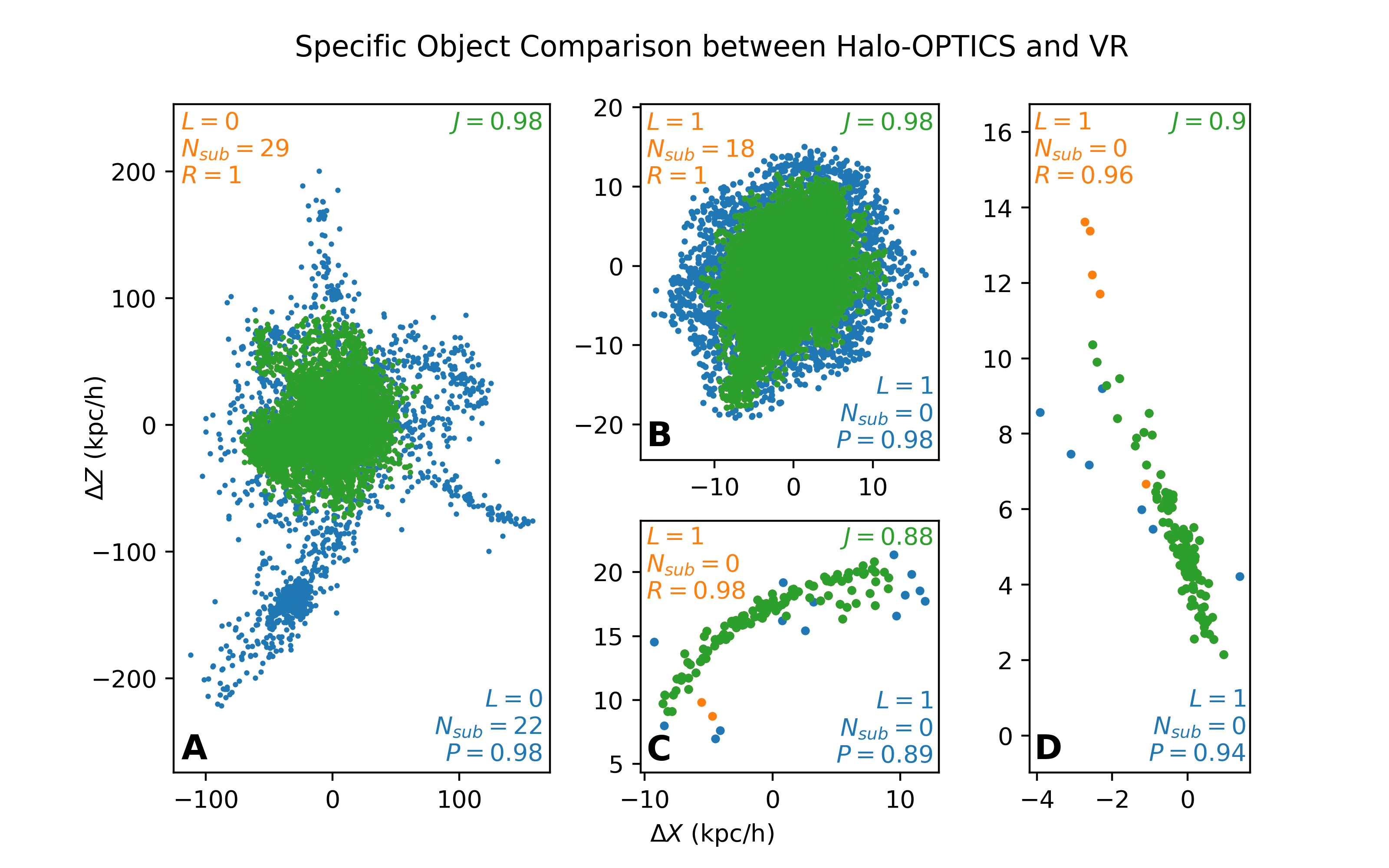}
    \vspace{-0.15cm}
    \caption{A series of 2D projections of a select few stellar clusters from the MW02 galaxy that have been found by both \code{Halo-OPTICS} and \code{VELOCIraptor}. Particles coloured; blue belong exclusively to the clusters as predicted by \code{Halo-OPTICS}; orange belong exclusively to the clusters as predicted by \code{VELOCIraptor}; green belong to the intersection of the clusters from both codes. Panels A, B, C, and D depict the \code{Halo-OPTICS} cluster and its best-fitting \code{VELOCIraptor}  candidate for the; root stellar cluster that surrounds the MW02 galaxy barycentre (large pink valley in top panel of Fig. \ref{fig:ReachabilityPlot}); the inner stellar halo of MW02 (partially shown in purple at the left-most edge of the middle panel of Fig. \ref{fig:ReachabilityPlot}); and two streams nearby the inner stellar halo (not explicitly shown in colour within Fig. \ref{fig:ReachabilityPlot} as they reside towards the left edge of the inner stellar halo's ordered list). The coordinate system of each panel is the same and is centred on the inner halo's barycentre. Annotated in orange in the upper left corner of each panel is the hierarchy level, number of substructures, and recovery of the cluster as predicted by \code{VELOCIraptor}. Similarly, annotated in blue in the lower right corner of each panel is the hierarchy level, number of substructures and purity of the cluster as predicted by \code{Halo-OPTICS}. Annotated in green in the upper right corner is the (maximum) Jaccard index of the two cluster representations.}
    \label{fig:specificcomparison}
\end{figure*}

Fig. \ref{fig:specificcomparison} illustrates the 2D projections of a select few clusters produced by \code{Halo-OPTICS} and their best-fitting \code{VELOCIraptor} counterparts. The panels therein indicate the particles attributed to each cluster by only \code{Halo-OPTICS} (in blue), only \code{VELOCIraptor} (in orange), and by both codes (in green). Various information about the cluster representations from the codes are annotated within each panel. We see that \code{Halo-OPTICS} provides a strong match to the predictions made from \code{VELOCIraptor} with high recovery, purity, and Jaccard index between each of the representations.

In panel A, \code{Halo-OPTICS} has retrieved extended stellar components -- that are likely of kinematic interest -- associated with the galaxy's surroundings that \code{VELOCIraptor} has not attributed to the galaxy. The inner halo from \code{Halo-OPTICS} depicted in panel B also over extends that from \code{VELOCIraptor}. Both of these over extensions, particularly the former, are largely resultant from the differences between the \code{OPTICS} and \code{FOF} algorithms -- i.e. \code{OPTICS} is not affected by noisy inter-particle spacing as it detects density fluctuations at the resolution of $N_{\rm min}$ points. The over extension seen in panel B is also due to the differences between the featured structure's spatial and phase-space densities -- a disparity that can likely be mitigated with the inclusion of particle kinematics into the \code{Halo-OPTICS} distance metric.

Panels C and D show that \code{Halo-OPTICS} does remarkably well in retrieving and matching these streams without the knowledge of particle kinematics. Given this knowledge, \code{Halo-OPTICS} could provide better quality matches to these streams than it does in the application we present here and potentially find some associated particles that \code{VELOCIraptor} does not.

We note that \code{Halo-OPTICS} does not find any of the substructures contained within the inner halo of the MW02 galaxy \footnote{This is by definition since we choose the inner halo to be the largest leaf cluster in the hierarchy.}. However, it should also be expected that \code{Halo-OPTICS} would do better in resolving these substructures with the knowledge of kinematics. Not indicated in Fig. \ref{fig:specificcomparison} is that of the $22$ sub structures found within the \code{Halo-OPTICS} root cluster depicted in panel A, $12$ are contained exclusively within the best-fitting cluster found by \code{VELOCIraptor} -- which contains $29$. By accounting for the known substructures in panels B, C, and D (and sub-substructures therein), we can deduce that there are precisely $8$ not-visualised substructures from \code{VELOCIraptor} within this region, and $8$ from \code{Halo-OPTICS}. These substructures are mostly the same between the codes however there is disagreement between the codes, namely the grouping shown in the lower panel of Fig. \ref{fig:ReachabilityPlot} and the inset panel of Fig. \ref{fig:RDClusters}. This grouping implies that there must be at least one other cluster found by \code{Halo-OPTICS} that is in dispute with those found by \code{VELOCIraptor}. Such clusters are likely to be clustered spatially but not kinematically, and given the same phase-space information as \code{VELOCIraptor}, we expect that \code{Halo-OPTICS} will find these clusters to be insignificant.

As mentioned in Sec. \ref{subsec:optics}, a major drawback to \code{OPTICS} -- and by extension \code{Halo-OPTICS} -- is that it is computationally demanding. For example, to complete a clustering run over the MW02 galaxy's stellar particles ($209834$ particles), \code{Halo-OPTICS} takes $\sim10$ minutes to create the reachability plot and then $\sim7$ seconds to extract clusters from that. For \code{Halo-OPTICS} to complete a clustering run over the dark matter particles ($2441561$ particles) within the MW02 galaxy, the runtime is $\sim5$ hours to create the reachability plot and then $\sim1$ minute to extract the clusters therein. In comparison, running \code{VELOCIraptor} over the MW02 galaxy's stellar particles only takes $\sim4.4$ seconds to search for substructure and $\sim25$ seconds to get the 6D\code{FOF} haloes. For the dark matter particles, \code{VELOCIraptor} takes $\sim16$ seconds to search for substructure and $\sim37$ seconds to get the 6D\code{FOF} haloes.

These runtime discrepancies are partially due to our naive implementation of \code{Halo-OPTICS} being run with \code{Python3} through a single core on an Intel Xeon E5-2698 v4 processor, whereas \code{VELOCIraptor} is a ready compiled program written in \code{C++11} with >= -O2 optimisation using gcc that in this instance used a single core and single MPI on an Intel i7 vPro processor. However, the largest runtime set back for \code{Halo-OPTICS} comes from the fact that not only does its nearest-neighbour radial search need to be much larger than the corresponding \code{FOF} nearest-neighbour radial search, but \code{Halo-OPTICS} also needs to return the exact distances of each neighbour during this search whereas \code{VELOCIraptor} does not. It should also be noted that as our implementation currently exists, \code{Halo-OPTICS} has no parallelisation capabilities and only uses optimised vectorised functions whereas \code{VELOCIraptor} has massively parallel capabilities. Using \code{VELOCIraptor} in this way can dramatically reduce the overall runtimes. For example, by allowing \code{VELOCIraptor} to use $8$ threads on the same dark matter particles as above, the substructure search time reduces to $\sim13$ seconds and the 6D\code{FOF} search reduces to $\sim26$ seconds.

\subsection{Inside the High Resolution Zone} \label{subsec:highresolutionzone}
\begin{figure}
\includegraphics[trim={5.5mm 5mm 14mm 18mm}, clip, width=\columnwidth]{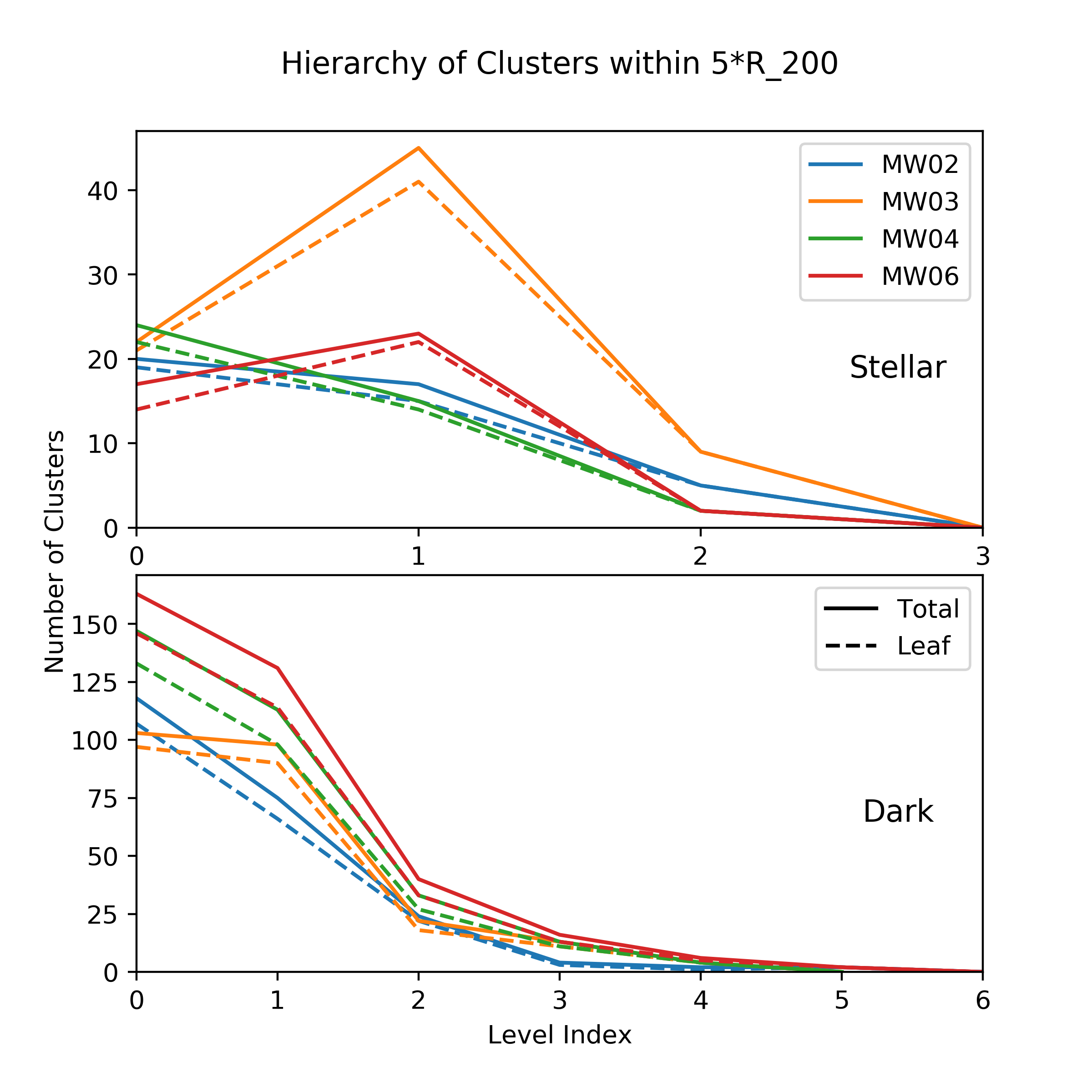}
\vspace{-0.5cm}
\caption{Stellar (top) and dark matter (bottom) cluster hierarchies for each galaxy. Levels 0--6 indicate the root to leaf layers of the hierarchy respectively. The solid and dashed lines illustrate the total number of clusters and the number of leaf clusters at each level respectively. The total number of leaf stellar clusters is $39$, $71$, $38$, and $38$ for the MW02, MW03, MW04, and MW06 respectively. Similarly, the total number of leaf dark matter clusters is $201$, $220$, $273$, and $313$ for the MW02, MW03, MW04, and MW06 respectively.}
\label{fig:hierarchy}
\end{figure}

Each synthetic halo has a virial mass ($M_{200}$) of approximately $2\times10^{12}\ \rm{M_\odot/h}$ \citep[details of each halo may be found in Table 1 of][]{Power2016}. The proportion of $M_{200}$ made up of stellar and dark matter particles also remains similar between haloes, however the total number of each type of particle within each galaxy snapshot does vary. The stellar and dark matter particles extend much further than $R_{200}$ and as such the size and complexity of the reachability plot varies as well. To reduce this variability between haloes we now take only those clusters whose barycentres are contained within $5R_{200} \approx 1\ \rm{Mpc/h}$ from the barycentre of the inner halo of each galaxy. This radial cut is chosen as it represents the approximate boundary of the high resolution regions within each of the cosmological zoom simulations that contain each MW type galaxy. Fig. \ref{fig:hierarchy} demonstrates the cluster hierarchy within this region of each of the galaxies for both the stellar and dark matter particles.

From Fig. \ref{fig:hierarchy} we see that within $5R_{200}$ each galaxy's hierarchy of clusters is similar. The number of clusters defined at the root level (level 0) is solely dependent on the \code{OPTICS} parameter $\epsilon$. By increasing $\epsilon$, the hierarchy will deepen overall and narrow at the zeroth level until eventually the zeroth level will only contain a single {\it cluster} -- the entire data set. To some extent the shape of the hierarchy at each particular level will be influenced by the choice of $\epsilon$, although since we use a rigorous definition for $\epsilon$ in our application of \code{Halo-OPTICS} the hierarchy shape between galaxies is meaningful.

Despite being born from the same simulation regime, the distinct hierarchy of stellar substructure within MW03 is seen in Fig. \ref{fig:hierarchy}, where the peak is noticeably larger in magnitude than the other galaxies. The stellar component of MW03 appears to exhibit the most galaxy-galaxy variation as seen in Figure 8 of \citet{Power2016}. Compared to the other galaxies the density of MW03 is large for small radii, drops for radii $\sim 0.1R_{200}$, and also features considerable spikes in density for radii approaching $\sim R_{200}$. The significant differences in stellar density within this region are certainly the reason for the large number of clusters at the first level of MW03's stellar hierarchy as these clusters will have a lower background density and therefore be in contrast with it more so than for the other MW type galaxies.

\section{Discussion} \label{sec:discussion}
We have demonstrated that the \code{Halo-OPTICS} algorithm is a powerful tool to be used for the global identification of all meaningful clusters of a data set containing at least $N_{\rm{min}}$ data points as well as the hierarchy within which they are embedded. When applied to the physical clustering of particles, the ambiguity of a meaningful metric disappears and the output becomes particularly robust when compared to other algorithms that operate under the same metric. We applied \code{Halo-OPTICS} to the 3D positions of stellar and dark matter particles from four MW type galaxies produced through a set of cosmological zoom simulations. We used \code{Halo-OPTICS} to detect and extract the significant clusters from these galaxies. We compared the output with \code{VELOCIraptor} before then analysing the hierarchy of clusters that are situated within a radius of $5R_{200}$ from their corresponding system's galactic centre.

Through our comparison with \code{VELOCIraptor} in Sec. \ref{subsec:comparison}, we have demonstrated that \code{Halo-OPTICS} retrieves the more significant clusters while electing to ignore those clusters whose density does not appreciably differ from their surroundings. Furthermore, this comparison indicates the power of adaptive hierarchical clustering algorithms such as \code{Halo-OPTICS} as it is able to uncover many clusters from only the 3D particle positions that \code{VELOCIraptor} had identified by using both particle positions and kinematics \footnote{This is not to say that \code{VELOCIraptor} is not adaptive -- it is -- \code{VELOCIraptor} iteratively uses a 6D \code{FOF} algorithm by locally adapting its phase-space metric to further separate clusters from their surroundings. This typically means that as the algorithm searches for clusters deeper within the hierarchy the phase-space becomes more heavily weighted towards particle kinematics rather than particle positions.}. For \code{Halo-OPTICS} to achieve this, these clusters must still have a significant spatial density contrast with their background, however for a 3D \code{FOF} algorithm to do the same, some level of fine-tuning would be needed.

The depth and shape of the hierarchy are influenced by the \code{Halo-OPTICS} hyperparameters. Those from the original \code{OPTICS} algorithm, $\epsilon$ and $N_{\rm{min}}$, are respectively responsible for the extents of lowest density and smallest size the clusters can be. Changing the \code{Halo-OPTICS} input parameter $\Delta$ has similar affect as its less physical \code{OPTICS} counterpart, $\epsilon$. The additional extraction parameters exclusive to \code{Halo-OPTICS} -- $\rho_{\rm threshold}$, $f_{\rm reject}$, and $S_{\rm outlier}$ -- are responsible for the number of divisions in between the root and leaf levels and which particles belong to each level and each cluster. However, the largest contributor to hierarchy is of course the physics of the interactions between the particles themselves. Being cold, the dark matter easily clumps together to form deep hierarchies by $z = 0$. Between baryonic feedback effects and the relative sub-dominance of stellar particles within the region defined by a radius of $R_{200}$ as a whole \citep[refer to Table 1 of][]{Power2016}, the resultant stellar hierarchies are shallower than their CDM counterparts. However due to stellar particles being kinematically cold, we should expect the stellar hierarchy to deepen with the inclusion particle kinematics.

It is likely that the inclusion of extra localised information -- i.e. velocities, chemical abundances etc. -- into the metric will have the largest impact on cluster yields in the inner regions of galaxies where large numbers of particles are very spatially dense and neighbouring local spatial densities are indistinct. This metric augmentation could conceivably be implemented as a non-linear combination of spatial, kinematic, and metallicity variables that are each weighted by a factor relative to their local variation within the data. Alternatively, \code{Halo-OPTICS} could perform its nearest neighbour searching over the spatial dimensions -- preserving the maximum spatial scales defined by the overdensity factor $\Delta$ -- and then order points by a distance metric containing information about spatial, kinematic, and chemical variables -- although this may not be necessary due to the adaptive nature of \code{OPTICS}.

Modifying the metric in this way will provide the means for determining clusters more distinctly from their background so long as the metric only includes good indicators of clustered data. The reachability plot will in general change shape for any given cluster, though not so significantly that we should not expect our cluster extraction method to still recover all relevant clusters. However, the optimal \code{Halo-OPTICS} hyperparameters may be different in a higher dimensional metric from those used in conjunction with a 3D spatial one.

The root levels of the hierarchy of clusters in a particular astrophysical data set will likely stay consistent across various metrics. Although, the hierarchy may deepen with the addition of extra clustering indicators since our cluster extraction process will be able to retrieve additional low spatial density and kinematically/metallicity coherent substructures at the leaf levels. Likewise, we may reasonably expect that changing the metric in these ways will not adversely affect the more massive substructures -- nor will it resolve any new ones -- and that the effect of an improved metric will predominantly modify the proportion of the less massive substructures compared to those that are larger.

\section{Conclusions} \label{sec:conclusions}
We have shown \code{Halo-OPTICS} to be a robust cluster finder that is effective in determining a wide variety of cluster types, shapes, and sizes, even with a spatial distance metric as its only handle on localised information. Furthermore, we are satisfied that our extraction process is capable of determining these clusters without the need for supervised learning nor the restrictions of the more conventional extraction techniques. The ability for the \code{Halo-OPTICS} algorithm to retrieve the hierarchy of galaxies in this relatively fast and secure manner should pave the way for \code{Halo-OPTICS} to be used complementary to a more traditional structure finder such as \code{VELOCIraptor}, and as a simple and practical halo finder in astrophysics and its related fields. In a future work, we will extend \code{Halo-OPTICS} to use a multi-dimensional metric that is inclusive of extra localised information, such as particle kinematics and stellar metallicity. We also intend to build upon our extraction technique so that it incorporates more physical aspects of clusters such as particle boundedness. Among these changes, we leave the further optimisation and potential parallelisation of \code{Halo-OPTICS} for future work as well. These concepts, particularly the latter, present significant challenges due to the strongly sequential data access order that \code{OPTICS} makes use of.

\section*{Acknowledgements}
WHO gratefully acknowledges financial support through the Hunstead Student Support Scholarship from the Dick Hunstead Fund in the University of Sydney's School of Physics. This research benefitted from computation resources in the form of the Argus Virtual Research Desktop environment which was provided through the University of Sydney's Information and Communication Technologies and supported by the Sydney Informatics Hub.

\section*{Data Availability}
The data underlying this article may be made available on reasonable request to the corresponding author.

\bibliographystyle{mnras}
\bibliography{optics}




\bsp	
\label{lastpage}
\end{document}
